%% file: main.tex
\documentclass[11pt]{article}

\usepackage[english]{babel}

\usepackage[letterpaper,top=1in,bottom=1in,left=1in,right=1in]{geometry}

\usepackage{amsmath}
\usepackage{graphicx}
\usepackage[colorlinks=true, allcolors=blue]{hyperref}
\usepackage[title]{appendix}
\usepackage{cite}
\usepackage{hyperref}
\usepackage{xcolor}
\usepackage{orcidlink}                
\usepackage[affil-it,blocks]{authblk} 

\begin{document}

\pagenumbering{gobble}

\title{\bf 
  \LARGE{Readout and PID using AIML for SoLID High Background Cherenkov Detectors}
\vspace{0.3in}

\large 

\author{Zhiwen Zhao}
\author{Bishnu Karki}
\author{Bo Yu}
\author{Andrew Smith}
\author{Gary Swift}
\author{Simon Gorbaty}
\author{Jingyi Zhou}
\author{Haiyan Gao}
\affil{Duke University}

\author{Benjamin Raydo}
\author{Alexandre Camsonne}
\author{Kishansingh Rajput}
\affil{Thomas Jefferson National Accelerator Facility}

\author{Marco Contalbrigo}
\author{Roberto Malaguti}
\affil{INFN Sezione di Ferrara, Italy}

\vspace{0.2in}

}

\date{} 
\maketitle

\include{abstract.tex}

\newpage

\clearpage
\tableofcontents
\setcounter{page}{1}
\pagenumbering{arabic}

\include{detector.tex}

\include{design.tex}

\include{test.tex}

\include{aiml.tex}

\include{sum.tex}

\bibliographystyle{unsrt}
\bibliography{ref,ref_aiml}

\end{document}

%% file: abstract.tex
\begin{abstract}
We present the development of readout electronics and artificial-intelligence-based particle-identification methods for the SoLID Cherenkov detectors at Jefferson Lab. To operate in the high-rate, high-background SoLID environment, we designed a MAROC sum readout system for multianode photomultiplier tubes that provides simultaneous pixel, quadrant-sum, and total-sum signals. Bench studies show that the system can sustain rates at or above those expected for SoLID while maintaining acceptable pedestal behavior and signal linearity. Using realistic Geant4 simulations for the heavy-gas Cherenkov detector, we then investigate $\pi/K$ separation with beam-related background. A simple photoelectron-counting cut is insufficient under these conditions, whereas multilayer perceptron models trained on PMT, quad, and pixel readout data perform substantially better. The quad and pixel readout schemes achieve pion and kaon efficiencies above 90\% and clearly outperform PMT-only readout. These results demonstrate that the combination of high-rate MAROC sum electronics and AIML-based pattern recognition provides a practical path toward robust SoLID Cherenkov PID.
\end{abstract}

%% file: detector.tex
\section{Cherenkov detectors of SoLID}

At the QCD intensity frontier of studying the nucleon structure and the strong force, high luminosity experiments and large acceptance detectors are crucial to collect valuable data from reactions with small cross sections and broad kinematic coverages. The Solenoidal Large Intensity Device (SoLID)~\cite{SoLID,solid_wp2022} is a new experimental apparatus proposed for Hall A at the Thomas Jefferson National Accelerator Facility (JLab). SoLID will combine large acceptance with the capability to handle high data rates at high luminosity to enable a slate of high-impact physics experiments. These include 3D imaging of the nucleon to probe the transverse momentum dependent parton distribution functions (TMD-PDFs, or just TMDs) through semi-inclusive deep inelastic scattering (SIDIS) and generalized parton distributions (GPDs) via various exclusive processes, testing the Standard Model and search for new physics using Parity Violating Deep Inelastic Scattering (PVDIS), and probing the origin of the proton mass via measurements of near-threshold production of the $J/\psi$ meson. 

\begin{figure}[!ht]
  \begin{center}
    \includegraphics[width=0.9\textwidth]{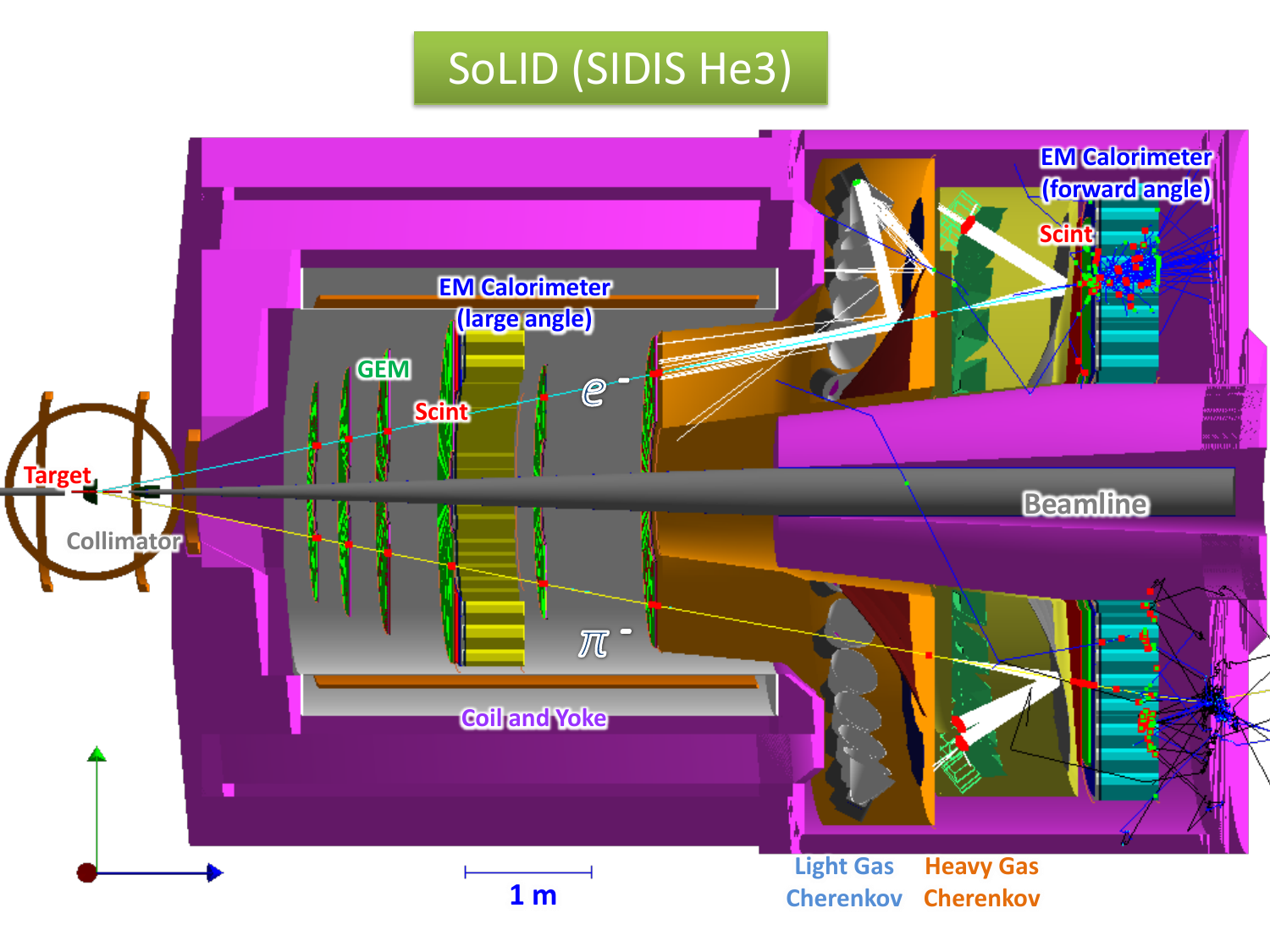}
    \caption{SoLID open geometry setup in Geant4 simulation.}
    \label{fig:setup}
  \end{center}
\end{figure}

The SoLID's open geometry setup for SIDIS measurements will allow for detecting electrons and charged pions in coincidence produced by 8.8 and 11 GeV electron beams on a polarized $^3He$ target. It covers roughly 8 to 24 degree polar angle and a full azimuthal angular range. To operate SoLID at the designed $10^{37}/cm^2/s$ luminosity for SIDIS measurements, we rely on recent developments in detectors, readout electronics, data acquisitions, and computing technologies. 

Two Cherenkov detectors in SoLID are used for particle identification (PID), a light gas Cherenkov (LGC) for $e/\pi$ separation and a heavy gas Cherenkov (HGC) for $\pi/K$ separation. Both have similar design, except LGC uses 1 atm CO2 gas while HGC uses 1.7 atm $C_4F_8$ gas. HGC has lower Cherenkov threshold of about 8 MeV for electrons than LGC's 16 MeV and thus higher background. We will focus on HGC in this study as it is more challenging.

\begin{figure}[ht!]
\vspace*{-0.1in}
\centering
\includegraphics[width=0.45\textwidth]{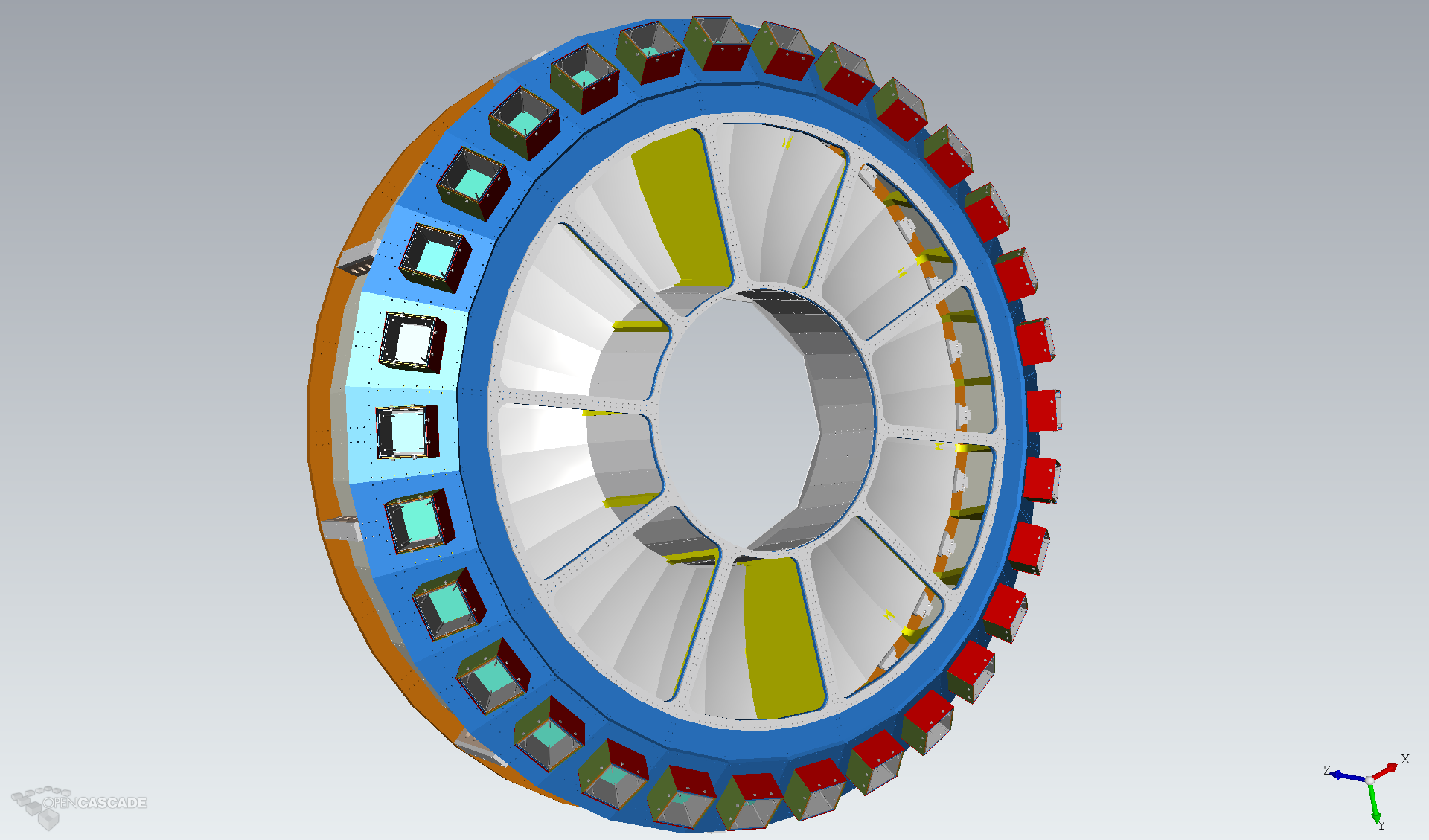}
\includegraphics[width=0.45\textwidth]{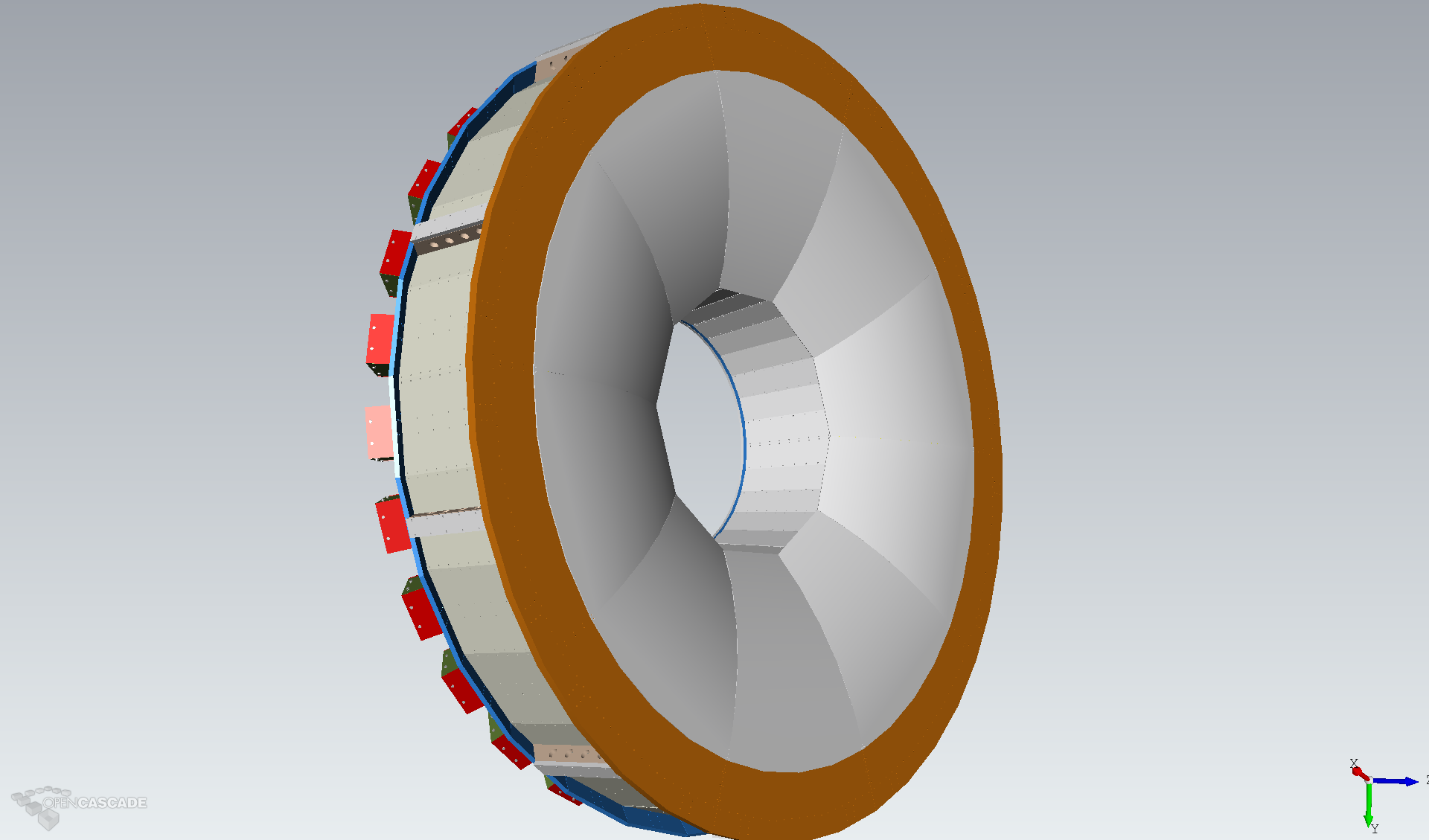}
\includegraphics[width=0.5\textwidth]{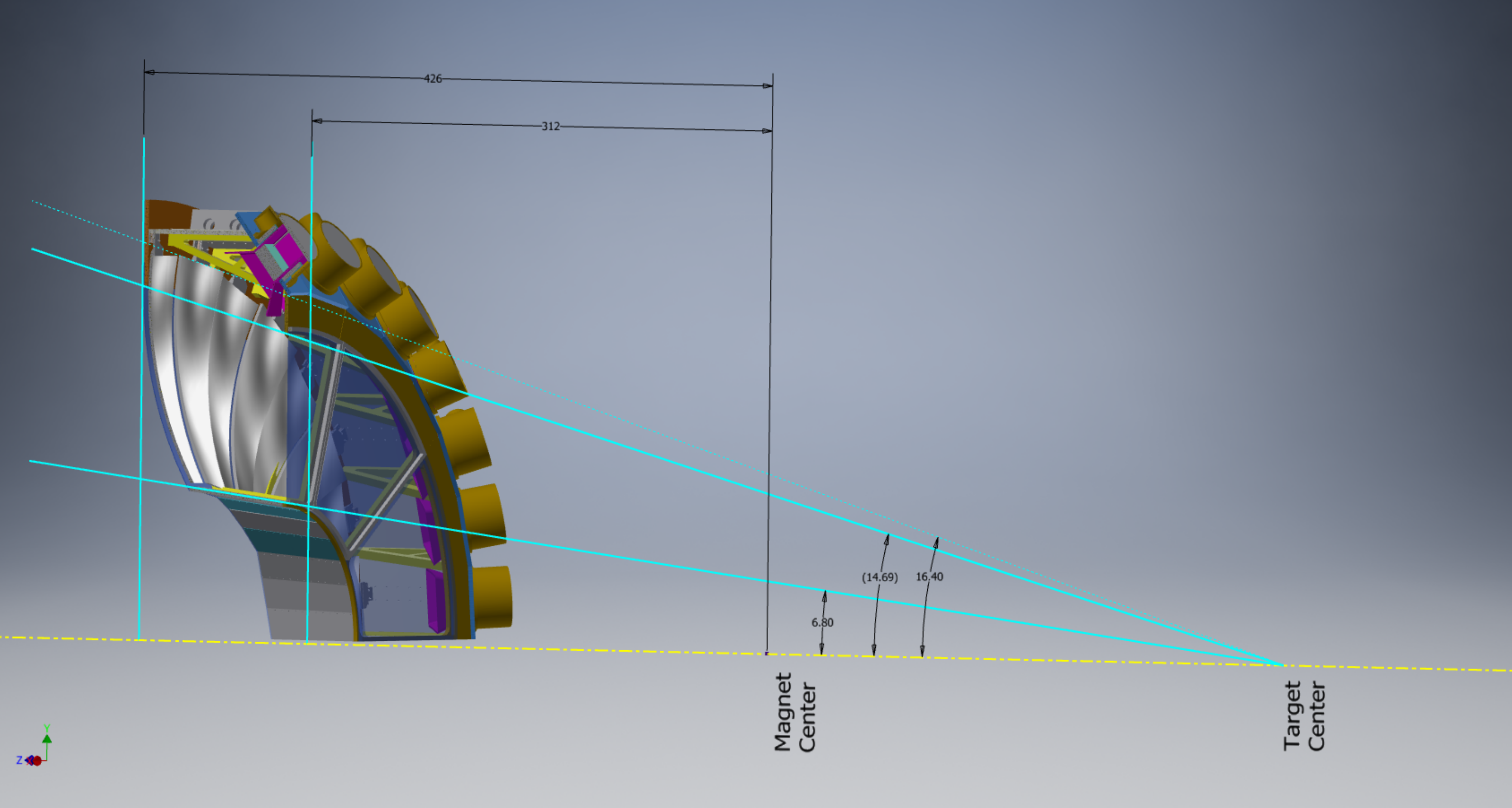}
\caption{Engineering design of HGC. The top left CAD picture shows the front view without front thin windows and the top right CAD picture shows the back view. The bottom CAD picture shows a few sectors in a cut view where the PMT array with light collection cone facing the mirror is visible.
}
\label{fig:hgc_structure}
\end{figure}

The HGC engineering design is shown in Fig.~\ref{fig:hgc_structure}. It covers the polar angle from 8 to 15 deg with 30 sectors in the  azimuthal direction. Each sector uses one spherical mirror to focus Cherenkov lights emitted along the 1 m gas length to a photonsensor plane. Our choice of photonsensors is multianode photomultipliers (MAPMT) H12700-03 from Hamamatsu. They are coated with a p-terphenyl wavelength shifter (WLS) to enhance the efficiency of UV light detection to have quantum efficiency about 30\% at 200 nm which matches the gas Cherenkov light spectrum well. The MAPMT has a square shape (5x5 cm). 16 of them are arranged in a 4x4 array as shown in the top photo of Fig.~\ref{fig:maroc_sum} as the photosensor plane in one sector. Each MAPMT has 64 pixels in an 8x8 array and each pixel (6x6 mm) has the sensitivity of detecting a single photon. The SoLID Cherenkov photon sensor choice is very different from traditional Cherenkov detectors using regular large size PMTs. To take advantages of the added spatial information of detected Cherenkov photons, special readout electronics are needed to be designed and tested, which is one of the main topics of this study.

In HGC, charged pions will emit enough Cherenkov lights when their momenta are higher than 2.5 GeV. It can be separated from charged kaons which won't emit Cherenkov lights until their momenta are higher than 7.5 GeV. However, at the $10^{37}/cm^2/s$ luminosity, there are a lot of lower energy electrons producing Cherenkov photons as background. It could be knock-on electrons from the HGC's thin (1 mm) Aluminum window by a pion or kaon passing through. And it could be electrons generated before HGC within the same time window (50 ns) when a pion or kaon passing through. This potentially makes the spatial information play an important role in PID besides only using the total number of photons traditionally. A more sophisticated PID algorithm is therefore needed to include the spatial information and Artificial Intelligence using machine learning (AIML) is a natural choice to solve such a pattern recognition problem, which is the other main topics of this study.

%% file: design.tex
\section{Readout design}

The HGC rate is expected to be as high as 200 kHz/pixel or 4 MHz/PMT during the experiment based on our Geant4 simulation. We need to have compatible readout electronics for MAPMTs with the capability of handling such a high rate.

During the 2018 beam test~\cite{Peng_2022}, the charges from a quadrant of 16 pixels were simply summed into one output. Each MAPMT thus has 4 quad outputs to feed into flash ADC (FADC). It operated well in a high background rate environment similar to SoLID. However, to truly taking advantage all pixel information with the flexibility of sum readout, a special readout electronics is needed.

\begin{figure}[!h]
    \centering
	\includegraphics[width=0.9\textwidth]{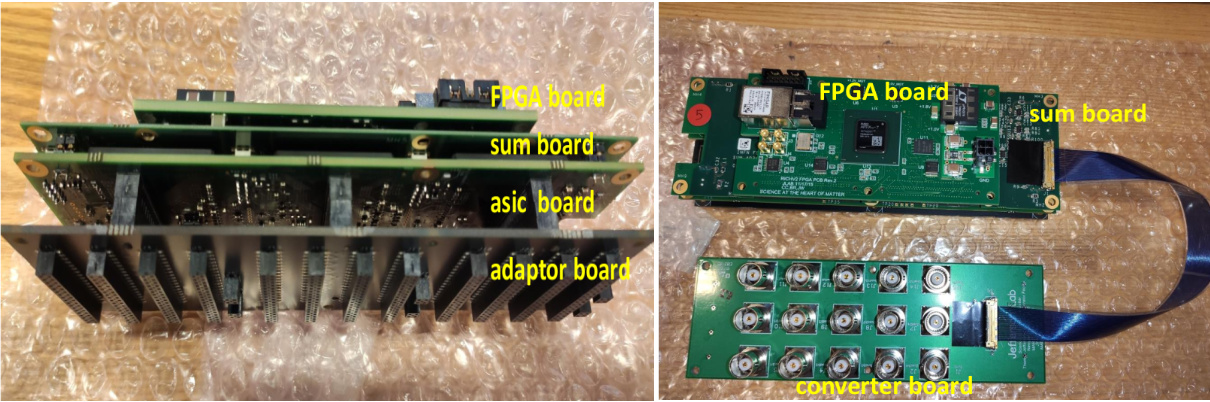}
\includegraphics[width=0.5\textwidth]{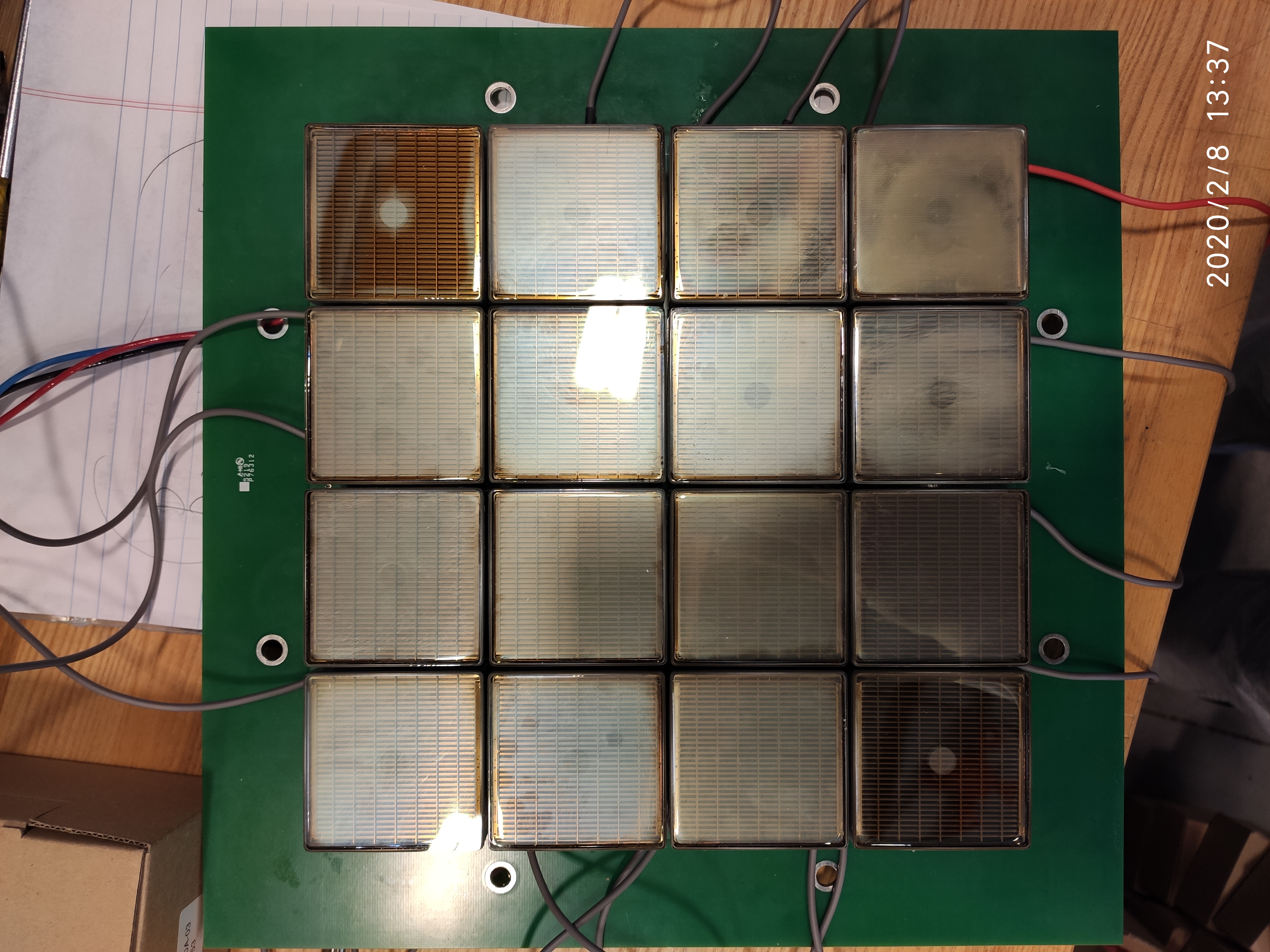} 
\includegraphics[width=0.4\textwidth]{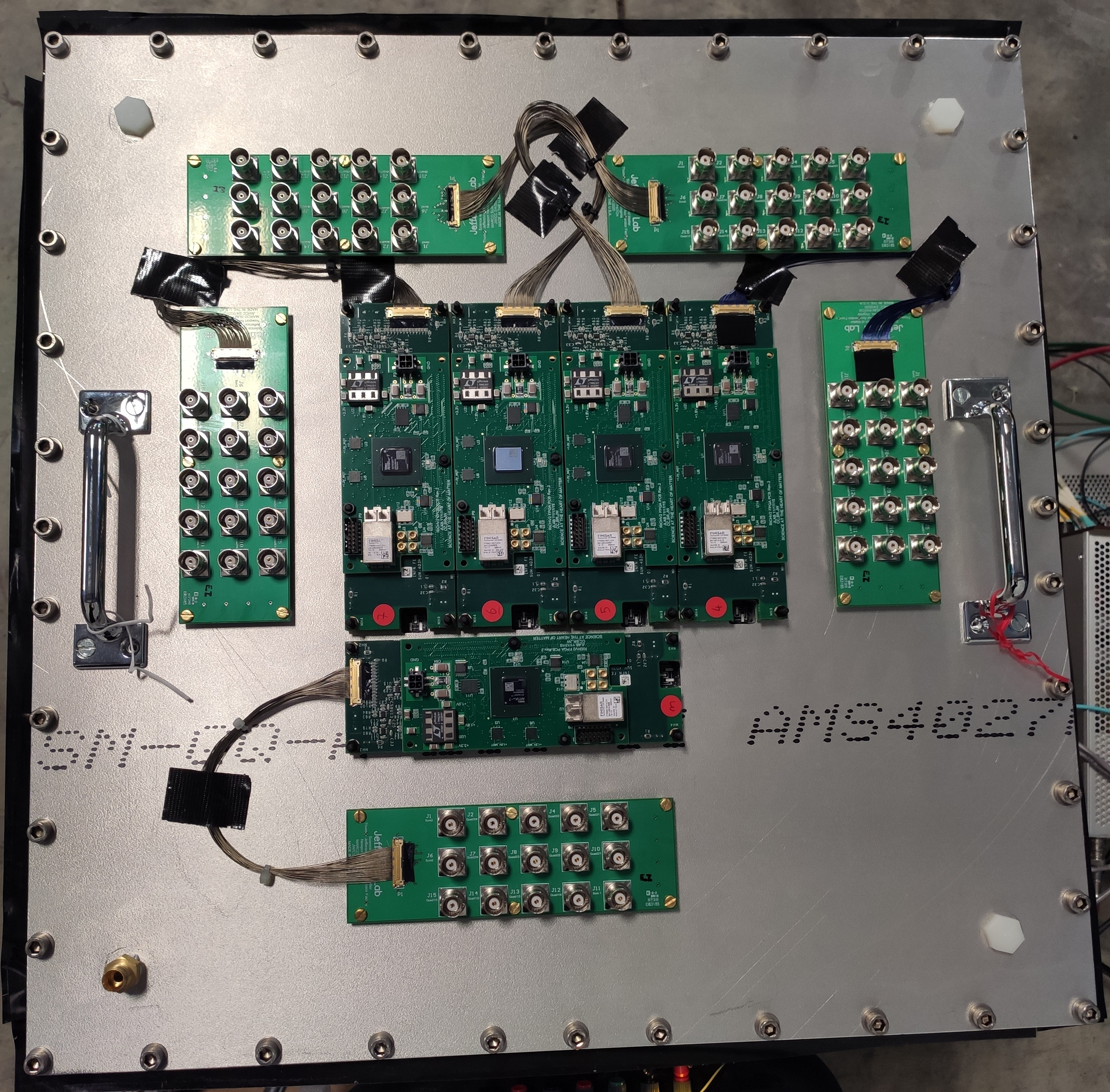}   
	\caption{(Top) photos of the MAROC sum readout system. It includes an adapter board, an ASIC board, a sum board, a FPGA board and a converter board. The MAROC ASIC board is modified from its original version to accommodate the new sum board. a new converter board connects the new sum board output to regular BNC connectors through an I-PEX cable. (Bottom) front and back photos of the 4x4 MAPMT assembly with MAROC sum readout.}
    \label{fig:maroc_sum}
	\end{figure}

The CLAS12 RICH detector~\cite{clas12_rich_nim} in Hall B at JLab uses the same type MAPMTs and their readout electronics is based on Multi-Anode Readout Chip (MAROC)~\cite{Blin:2010tsa}. 
Their MAROC readout system includes an adapter board to hold up to 3 MAPMTs, an ASIC board hosting MAROC chips, and a FPGA board with optical fibers to communicate with the Subsystem Processor (SSP) module in a Data Acquisition (DAQ) VXS crate to record TDC signals from 192 pixels of 3 MAPMTs. For SoLID Cherenkov detectors, to collect pixel information and provide trigger with the sum of photons simultaneously, a new MAROC sum readout system was designed and manufactured in collaboration with the INFN Ferrara group and the JLab fast electronics group. By inserting a new sum board between a modified ASIC board and an FPGA board, charges in pixels are summed and separated into four quadrant sums and one total sum. A new converter board then transfers the sum signals to regular BNC connectors to be read by FADC. The top photos in Fig.~\ref{fig:maroc_sum} show the five boards mentioned above and how they are connected together to form the MAROC sum readout system. And the bottom photos show the front and back view of the 4x4 MAPMT assembly with MAROC sum readout which we built and used for testing.

\begin{figure}[!h]
    \centering
	\includegraphics[width=0.25\textwidth]{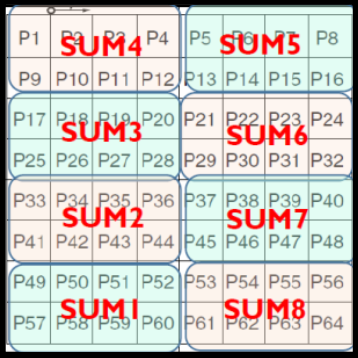}
		\includegraphics[width=0.7\textwidth]{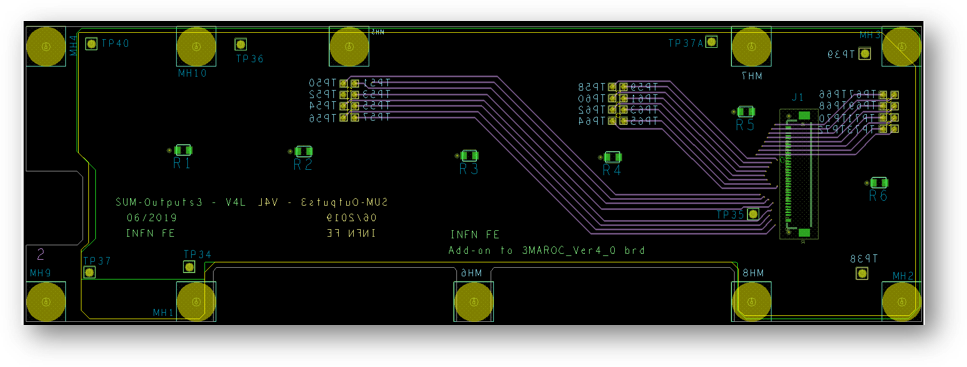}
	\caption{(Left) The schematic representation of the summing mechanism of MAROC sum board for H12700-03 MAPMT series with 64 pixels. The sum board sums group of 8 pixels. The SUM1 and SUM2 give the 1st quadrant signal, SUM3 and SUM4 give the 2nd quadrant, and so on. The total sum represents the sum of charge collected in all 64 pixels coming from 8 sums. (Right) The sum board electrical design.}
    \label{cher:quad_sum_mechanism}
	\end{figure}

The sum board provides analog sums of charges collected in groups of pixels. It first sums the charge collected in a group of 8 pixels to provide a total of 8 analog sums as shown in Fig.~\ref{cher:quad_sum_mechanism}. Then in the final output, for a quad signal, two of 8 sum signals are added. To be exact, Quad 1,2,3,4 are the sum of SUM 1 and SUM 2, SUM 3 and SUM 4, SUM 5 and SUM 6, SUM 7 and SUM 8, respectively. For the total sum of all pixels in a PMT, all 8 sums are added.

%% file: test.tex
\section{Readout test}

We tested the MAROC sum readout system to evaluate its features and performances and the results are presented in this section.

\subsection{Saturation study of sum signals}

\begin{figure}[!h]
    \centering
	\includegraphics[width=0.7\textwidth]{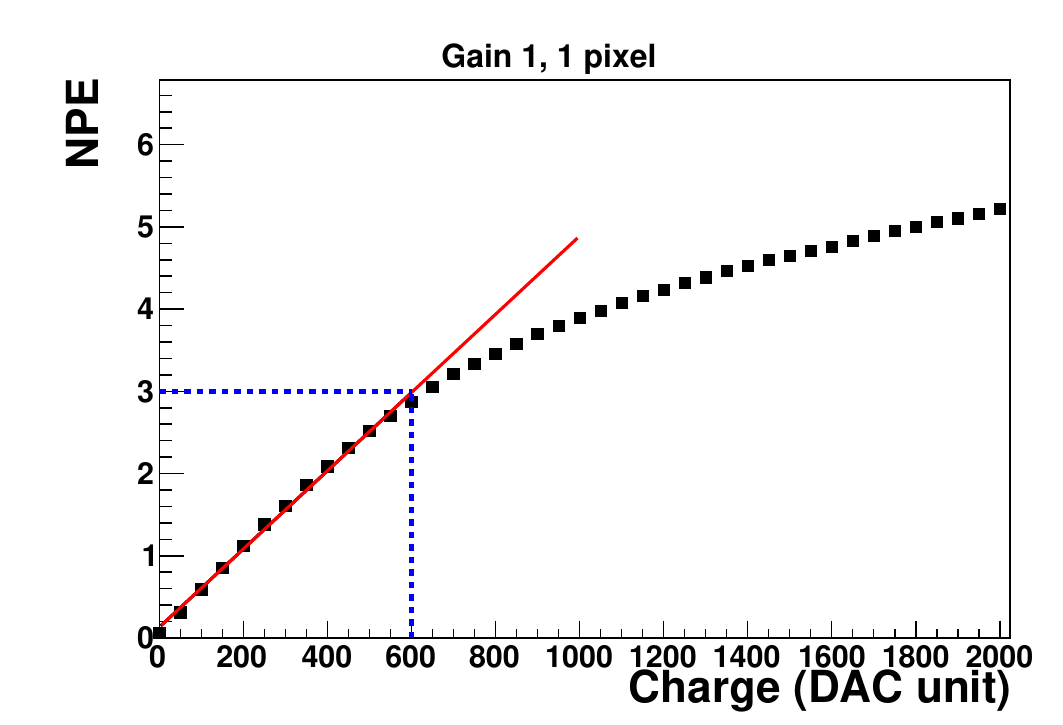}
	\caption{Saturation study of MAROC sum board with one pixel fired. The number of photoelectrons vs injected charge in DAC unit. The red line is a linear fit to the data. The maximum value of charge injected and the number of photoelectrons before saturation is shown by the dashed blue vertical and horizontal lines for MAPMT at 850 V.  }
       \label{fadc_to_npe}
	\end{figure}

To correctly measure number of photoelectrons (NPE), the MAROC analog sum output should be linear to the amount of charge injected. When the injected charge is too large, the MAROC sum starts to saturate and the output signal become nonlinear to injected charge. We carried out a study to determine the maximum number of photoelectrons that a single pixel, a quad, and the total sum signal can hold before saturation. For our test, we injected the known charge in a given pixel or group of pixels in an increasing step and recorded the MAROC sum output. Fig.~\ref{fadc_to_npe} shows the number of photoelectrons for different injected charges with only one pixel fired on a MAPMT at 850 V. The red line is a linear fit to the data. The blue vertical and horizontal dashed lines represent the threshold charge in the DAC unit and the number of photoelectrons respectively above which the sum signal saturates. The NPE values are calibrated through the measurement of the ADC mean position of a Single Photoelectron (SPE) distribution for a pixel can withstand before saturation. The saturation limit varies with the number of pixels fired in a group of pixels shown in Fig.~\ref{cher:quad_sum_mechanism}. Table~\ref{tab:kin} shows the maximum number of photoelectrons for an individual pixel, a quad, and total sum below which the MAROC sum is linear to the injected charge. We performed our study by applying different HV to MAPMT and found that the linear range decrease with an increase in HV and corresponding gain.

	\begin{table}[!h]
	\begin{center}
	\scalebox{1.0}{
  \begin{tabular}{|c|c|c|c|}
    \hline 
    {NPE} &\multicolumn{3}{c|}{Number of pixels fired}\\
    \hline
    HV (V) & 1 pixel & 16 pixels (Quad) & 64 pixels (Total Sum)\\
   \hline
   850  & 3 & 10 & 44 \\
   \hline
   900  & 2 & 7 & 29 \\
   \hline
   1000  & 1 & 4 & 17 \\
   \hline
  \end{tabular}}
  \end{center}
  \caption{Saturation limit for the MAROC sum board in terms of NPE for a single pixel, quad, and total sum signal at different high voltage for MAPMT. With the increment in HV applied to MAPMT the linear dynamic range gets reduced. }
  \label{tab:kin}
\end{table}

\subsection{Pixel TDC threshold study}
\label{tdc_threshold}

\begin{figure}[tb]
    \centering
	\includegraphics[width=0.7\textwidth]{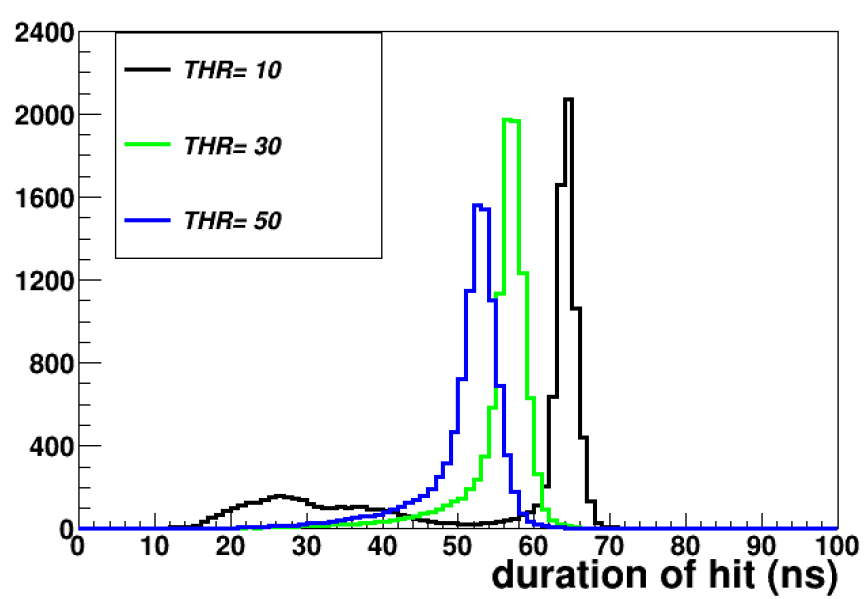}
	\caption{Duration of hit distribution from pixel readout of the MAROC sum board at 3 different values of thresholds 10, 30, and 50 DAC units above the average pedestal position, with MAPMT at 850 V.}
    \label{Threshold_scan}
	\end{figure}

\begin{figure}[h]
    \centering
	\includegraphics[width=0.6\textwidth]{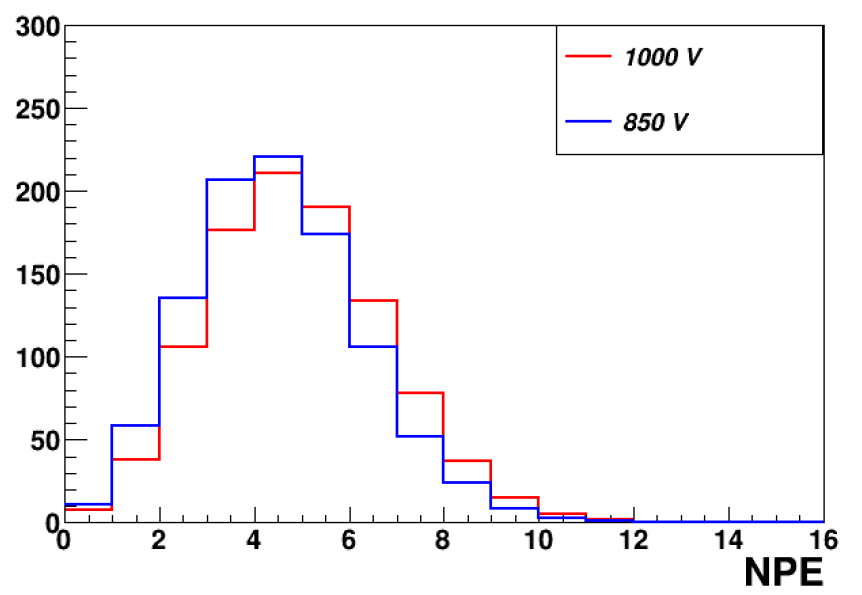}
	\caption{A number of photoelectrons (NPE) distribution with TDC threshold +30 DAC unit above pedestal for two identical settings. The NPE distribution with MAPMT at 1000V is represented by the red histogram while the blue histogram is for MAPMT at 850 V.}
	
    \label{cher:TDC_effi}
	\end{figure}

For MAROC pixel readout like in CLAS12 RICH, MAPMTs are usually operated at around 1000 V. But to have both pixel and ADC readout we found that one can get the benefit of a larger linear region while operating MAPMTs at lower voltages as shown in the previous section. To find the optimized value of the TDC threshold at 850 V, we studied the duration of a hit at different TDC thresholds. The duration of a hit is defined as the difference in time when the trailing and leading edge cross the TDC threshold and it depends on the amplitude of a signal. For cross-talks with small amplitudes, the duration of a hit is small compared to a real signal. Fig.~\ref{Threshold_scan} shows the duration of hit distribution for 3 typical TDC discriminator threshold values 10, 30, and 50 DAC units above the average pedestal value. At the lower threshold, we recorded a lot of lower amplitude noises and at the higher threshold, we may miss some real signals with small amplitudes. After these tests, the common discriminator threshold was chosen to be +30 DAC units above the average MAPMT pedestal position, a level that corresponds to a small fraction of the average SPE amplitude. To make sure that our choice of TDC threshold does not reject the SPE signal we compared the number of photoelectrons distributions for two identical runs with the same TDC threshold, +30 above the mean pedestal. The red and blue distributions in Fig.~\ref{cher:TDC_effi} correspond to a MAPMT at 1000 V and 850 V respectively. The mean of the NPE distribution at 850 V is about 95\% of the distribution at 1000 V. It suggests that MAPMT at 850 V, TDC threshold of +30 DAC units above the mean pedestal is sufficient to record the small-amplitude signal with minimal loss of efficiency.

\subsection{High rate bench test}

\begin{figure}[!h]
    \centering
	\includegraphics[width=0.7\textwidth]{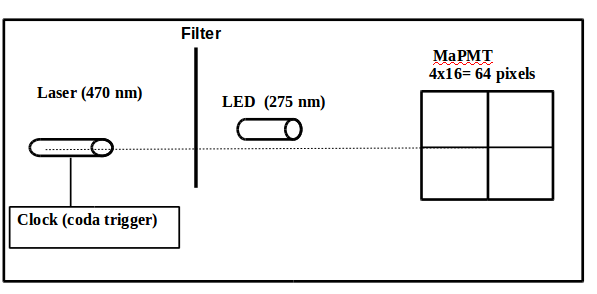}
\caption{Schematic layout of the bench test. The entire setup was placed inside the black box to minimize the background.}
    \label{cher:Schematic_layout}
	\end{figure}

One of our goal is to test the performance of MAROC sum electronics up to the maximum expected SoLID rates, 200 kHz/pixel for pixel readout or 4 MHz/PMT for sum readout. We designed a bench test with its schematic layout shown in Figure~\ref{cher:Schematic_layout}. 
Both a laser (470 nm) and a LED (275 nm) were used as light sources. The amount of lights from the LED was controlled by changing the applied DC voltage. A diffuser, labeled as filter, was placed between the laser source and MAPMT to choose different laser light intensities: strong, medium, and weak. For both light sources, we started from low light condition and increased lights slowly. Together they helped to reach the needed high rate. Photons from the LED and the laser were collected by a WLS coated Hamamatsu H12700-03 MAPMT with MAROC sum readout. MAROC output had two independent pieces of information: signal from 64 pixels and 5 sum signals including 4 quads of 16 pixels each and 1 sum of 64 pixels. An independent pulse generator was synchronized with the laser to control its pulsing and also as the trigger for the daq system.

\par
\begin{figure}[tb]
\centering
\includegraphics[width=0.7\textwidth]{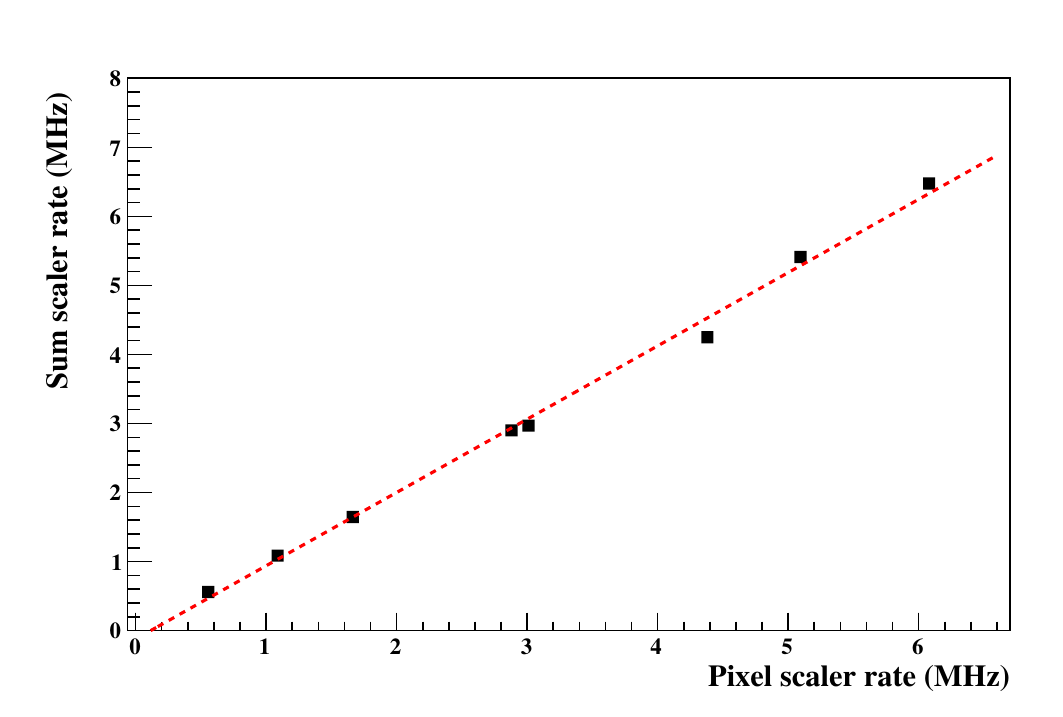}
\caption{Comparison between the sum and pixel scaler rates. The agreement between the two scaler rates are within 3\% for rates similar to that expected in SoLID. }
\label{cher:tdc-fadc-scaler}
\end{figure}

 
To ensure the desired rate is achieved, we measured both the sum scaler rate and single pixel scaler rate for every pixel at the same time. A single event may result in multiple pixel hits but a single hit in the sum signal. The sum scaler rate should match the pixel scaler rate calculated from single pixel scaler rates. The calculation is according to Equation~\ref{relation_pixel_sum}.
\begin{equation}
\label{relation_pixel_sum}
    \mbox{pixel scaler rate} = \frac{\mbox{average single pixel scaler rate} \times 64}{\mbox{average pixel occupancy}}
\end{equation}
where the average pixel scaler rate is an average among the 64 pixels and the average pixel occupancy is an average number of pixel hits for an event.
Fig.~\ref{cher:tdc-fadc-scaler} shows the agreement between the sum and pixel rates is within 3\%. We achieved the rate $\sim$ 6 MHz/PMT, larger than larger than 4 MHz/PMT expected in SoLID.	

\par
\begin{figure}[tb]
    \centering
	\includegraphics[width=0.45\textwidth]{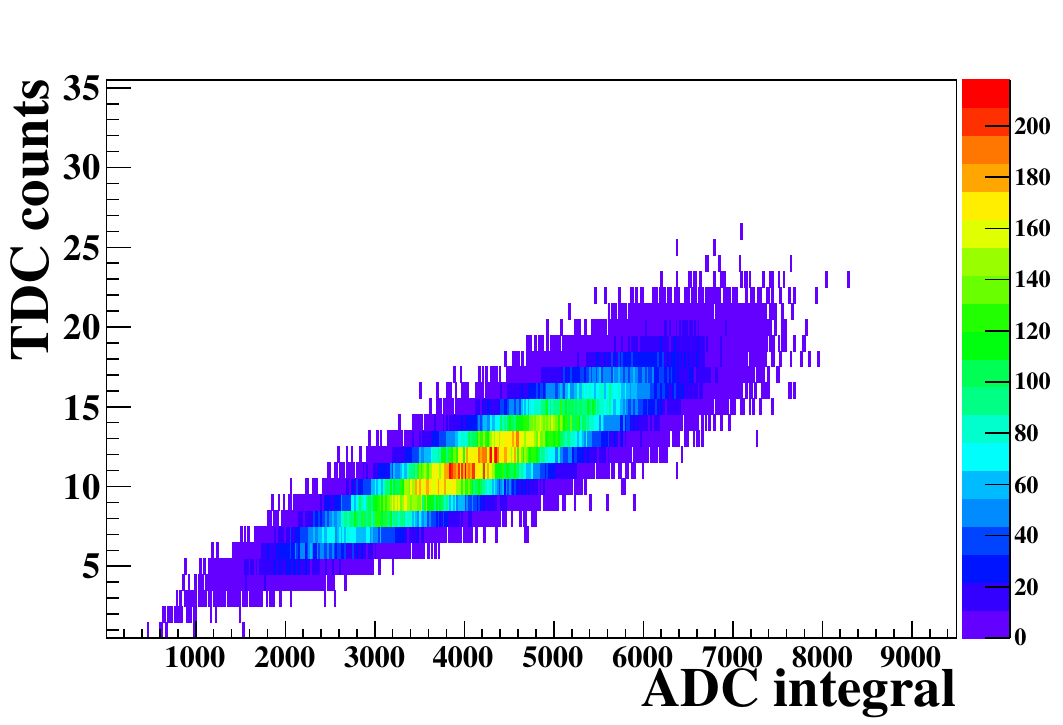}
	\includegraphics[width=0.45\textwidth]{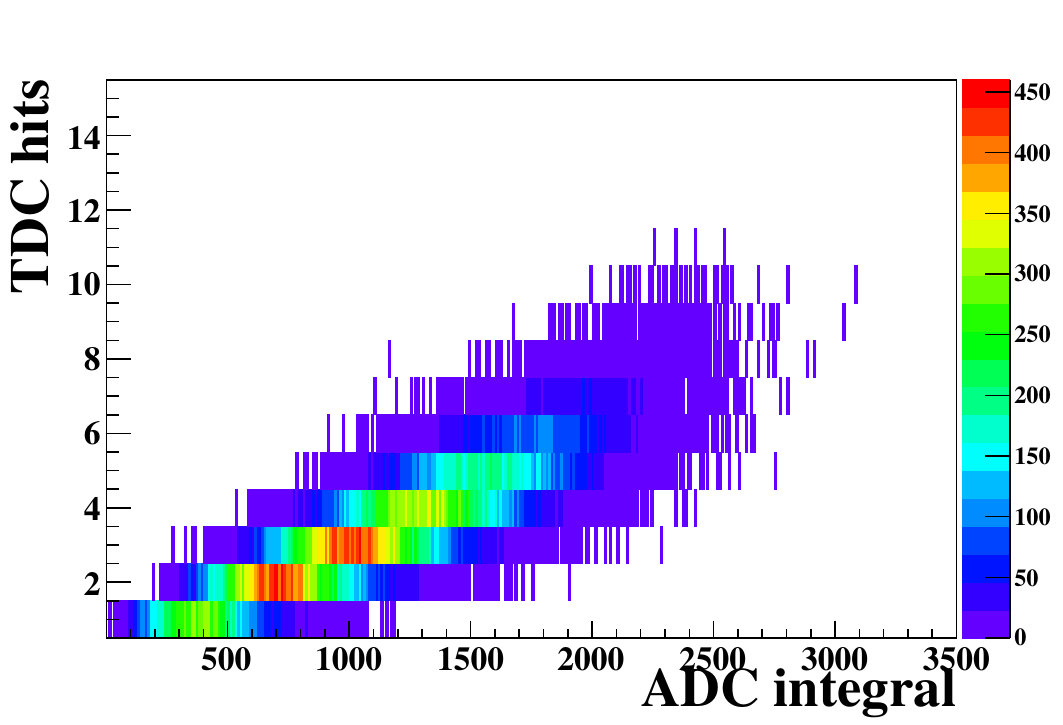}
	\caption{Left: TDC counts VS sum ADC integral. Right: TDC counts vs quad ADC integral. The linear correlation between the pixel readout by MAROC TDC and both the sum and quad readout by FADC is observed for the MAPMT with MAROC sum readout.}
    \label{cher:correaltion}
	\end{figure}

For each event, the independent information coming from pixel signal read out by MAROC TDC and quad and sum signals read out by FADC should represent the same event. If the electronics is performing well at high rate then one should expect the linear correlation between these two type of readouts. We verify the linear correlation between the TDC counts and ADC from integrating a FADC signal over its time window as shown in Fig.~\ref{cher:correaltion} up to the sum rate of $\sim$6 MHz/PMT. This demonstrates that summing electronics work as expected to collect the charge from pixels.

\subsection{Ring test}

In Cherenkov detectors, the Cherenkov ring is used for particle identification. To study the performance of MAROC sum electronics and the efficiency of our algorithms we analyzed the ring formed at MAPMT in presence of background comparable to SoLID running condition. On our setup the background was generated by LED operated by DC voltage. We had two setups to produce the rings, one using LEDs with ring shaped filters and the other using cosmic muons emitting real Cherenkov photons. We analyzed these rings for both pixel and sum readout using two algorithms:
\begin{itemize}
    \item Number of photoelectron cuts which is most straightforward
    \item Hough circular transform, the ring finder algorithm borrowed from computer vision studies
\end{itemize}
To test the performance our algorithm, we define the term accuracy for accepting signal and rejecting background as in Eq.~\ref{FOM1} and Eq.~\ref{FOM2} respectively. The ideal result would have both FOMs' as large as possible or at least larger than 90\%.

\begin{equation}
\begin{aligned}
    (Accuracy)_{signal} &= \frac{\text{number of accepted signal events }}{\text{total signal events}}\\
\end{aligned}
\label{FOM1}
\end{equation}
\begin{equation}
\begin{aligned}
    (Accuracy)_{background} &= \frac{\text{number of rejected background events}}{\text{total background events}} 
\end{aligned}
\label{FOM2}
\end{equation}

\subsubsection{LED Ring Test}

\begin{figure}[!h]
    \centering
	\includegraphics[width=0.55\textwidth]{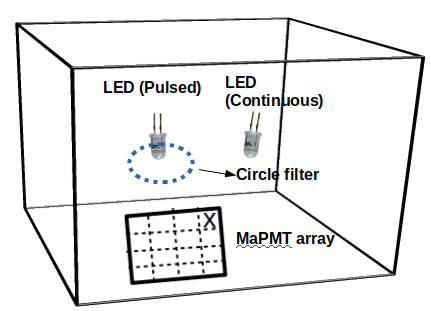}
\caption{Schematic layout for the LED ring test. For signals, lights from the pulsed LED was passed through the circular filter before it is collected in the MAPMTs array. The second LED operated with DC voltage provides the background. The entire setup was placed inside a black box to avoid any external light.}
    \label{cher:led_test_layout}
	\end{figure}

Two similar LEDs (275 nm) were enclosed inside a box. The pulsed light from the first LED was passed through the circular filter to form a ring near the center of the MAPMT array with a radius close to 8 pixels. The second LED was operated with DC voltage to provide continuous background. The background level from the second LED was controlled by adjusting the voltage of its DC power supply. Fig.~\ref{cher:led_test_layout} shows the schematic experimental setup to produce the LED rings. We collected the data at a different level of background up to almost twice as expected in the SoLID running condition. Position and ring size are fixed for the LED ring so they are relatively simple to identify from the background. But they still help us to test the performance of our readout and algorithm.

\begin{figure}[!h]
    \centering
	\includegraphics[width=1.1\textwidth]{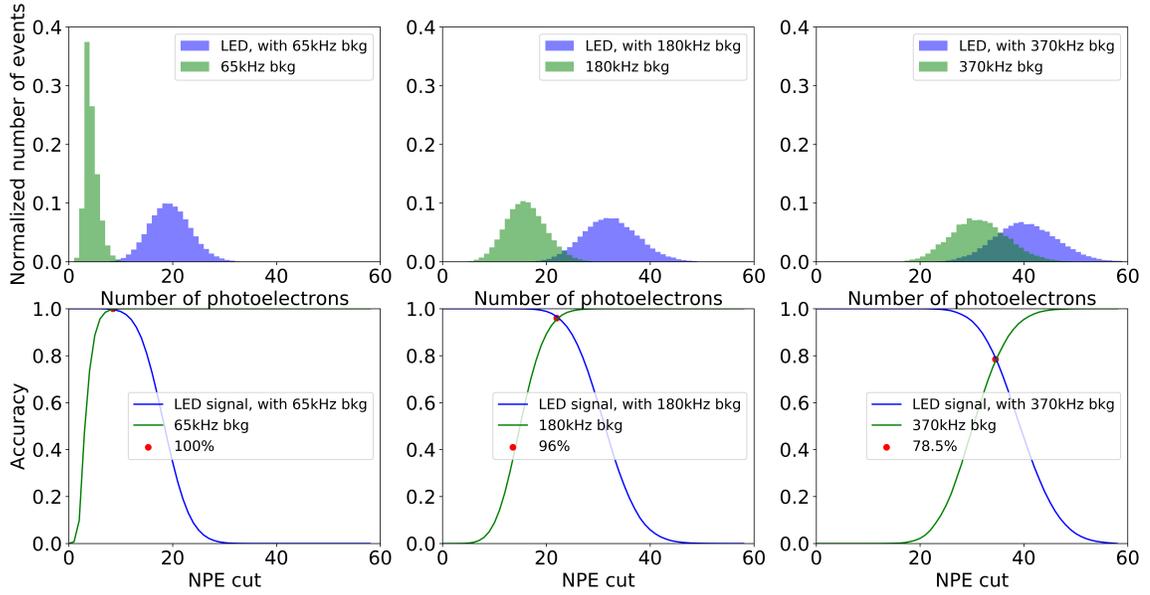}
\caption{Top: Number of photoelectron distribution for LED ring with pixel readout. The green and the purple histogram represents a number of photoelectrons with background only and LED signal in presence of background. The histograms are at the background rate of 65, 180, and 370 kHz/pixel from left to right. Bottom: The figure of merit for the number of photoelectrons cut at different levels of background.}
    \label{pixel_accuracy}
	\end{figure}

\begin{figure}[!h]
    \centering
	\includegraphics[width=0.5\textwidth]{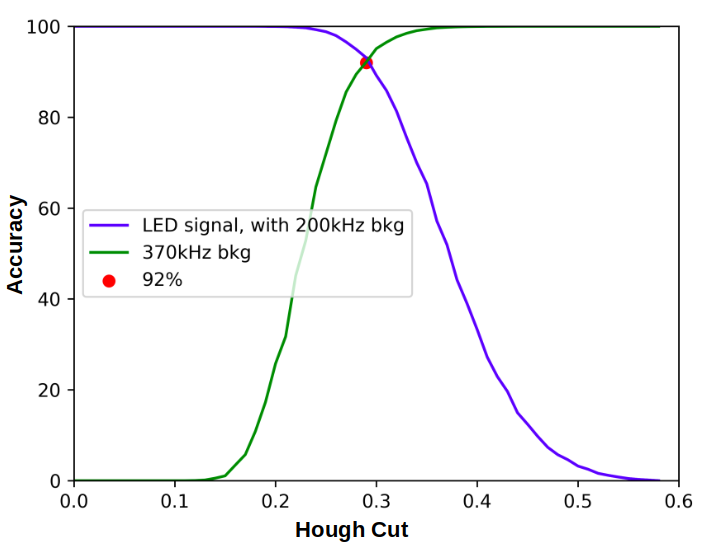}
\caption{The accuracy for the Hough transformation algorithm at background of 370kHz/pixel. At high background environment the Hough transformation performs well compared to number of photoelectron cut. }
    \label{hough_transform_pixel_370}
	\end{figure}

\begin{figure}[!h]
    \centering
	\includegraphics[width=1.0\textwidth]{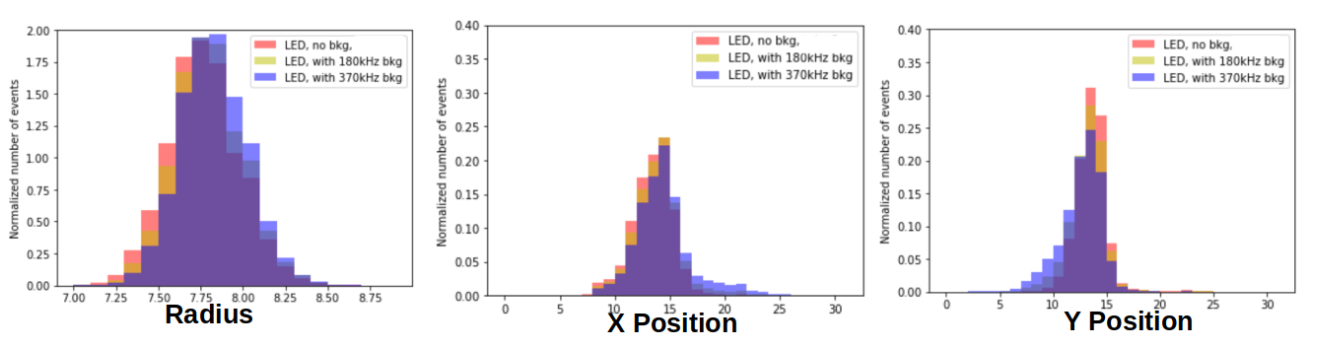}
\caption{Parameters of the recognized rings using the Hough transformation for pixel readout. The reconstructed radius and position of the ring are in agreement with the experimental setup.}
    \label{ring_parameters}
	\end{figure}

Fig.~\ref{pixel_accuracy} shows that at the lower background level, less than 200 kHz/pixel, the background distribution is well separated from the signal in the presence of background. At these lower background rates, the simple number of photoelectron cuts is sufficient enough to separate the signal from a background with better than 90\% accuracy. However, at a higher background rate, the background and signal (in presence of background) distribution are more overlapped and the number of photoelectron cut is not sufficient enough to reject the background from the signal. We then used the more sophisticated Hough transform algorithm for the high background test data. Fig.~\ref{hough_transform_pixel_370} demonstrates with Hough transformation, the accuracy is improved from 70\% to 92\% at the random background of 370 kHz/pixel. The parameters of the recognized rings were reconstructed using Hough transformation. Fig.~\ref{ring_parameters} shows the reconstructed parameters of ring: radius, X, and Y positions of the LED ring at various background rates. The reconstructed parameters are in good agreement with the experimental setup.

We also analyzed the sum readout results for the same LED ring data. The quad sum readout using Hough transformation method can provide similar separation with larger uncertainty than the pixel readout. It's much more difficult for total sum readout to use Hough transformation method because the ring only covers about 4 PMTs.

\subsubsection{Cosmic Ring Test}

\begin{figure}[!h]
    \centering
	\includegraphics[width=0.75\textwidth]{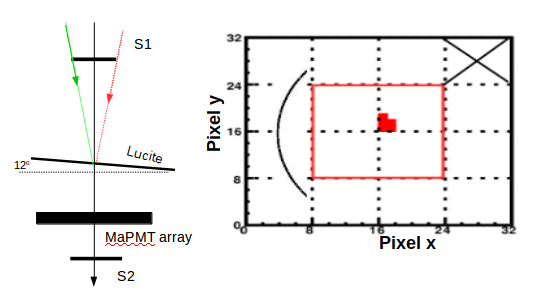}
\caption{Left: Schematic layout for the cosmic ring test. Two scintillator bars, one at the top and another at bottom of the PMT array form the trigger. The lucite block is tilted at an angle 12$^\circ$ with horizontal. The red and the green lines represent the extreme of scintillator acceptance. Due to the total internal reflection and the limited size of the MAPMT array, only about half of the triggered events form the partial ring. 
Right: The MAPMT array of dimension $\sim$ 20 x 20 cm is used to detect the Cherenkov ring. The cosmic muon can hit any of the central 4 PMTs represented by the red rectangular box. }
 \label{cher:cosmic_test_layout}
	\end{figure}

We also tested the Cherenkov rings produced using cosmic muons. Figure~\ref{cher:cosmic_test_layout} shows the schematic layout for the cosmic ring test. As a cosmic muon passes through a 4 cm lucite radiator with refractive index of 1.5, the Cherenkov photons are emitted then collected in the MAPMTs array. The opening angle of Cherenkov light (48.2$^{\circ}$) is larger than the critical angle of lucite for total internal reflection (41.8$^{\circ}$). As a consequence, the Cherenkov light emitted with vertical cosmic muon won't be able to pass through the lucite surface to reach to the MAPMT array. To take the advantage of vertical muon flux we tilted the lucite by 12$^{\circ}$. After tilting the lucite we were able to collect the Cherenkov light produced by the muon with an incident angle between 0 to 6$^{\circ}$ (angular acceptance of muon between the green and vertical black lines) while muon striking from another half (between the red and vertical black lines) were missed due to the total internal reflection and acceptance of the MAPMT array. The trigger was formed by two scintillator bars S1 and S2 in coincidence with about 5 cm x 5 cm overlapping area. The muon can hit anywhere within the 4 central MAPMTs shown by the red rectangle and form a cluster at the MAPMT array. The expected position of partial (about 35\%) Cherenkov ring and muon cluster is shown in Figure~\ref{cher:cosmic_test_layout} (right).

\begin{figure}[!h]
    \centering
	\includegraphics[width=1.1\textwidth]{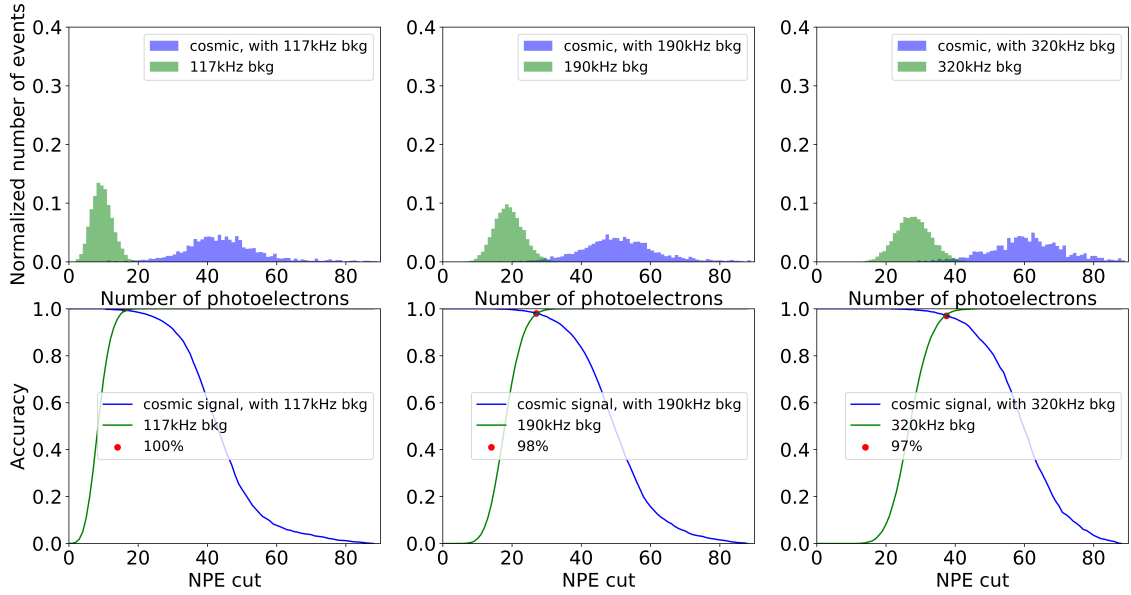}
\caption{Top: Number of photoelectron distribution for cosmic data with pixel readout. The green and the purple histogram represents a number of photoelectrons with background only and cosmic in presence of background. The histograms are at the background rate of 117, 190, and 320 kHz/pixel from left to right. Bottom: The figure of merit for the number of photoelectrons cut at different levels of background.}
    \label{pixel_accuracy_cosmic}
	\end{figure}

For the cosmic test setting, we computed the expected number of photoelectrons $\sim$30 with the MAPMT quantum efficiency, partial ring due to total internal reflection, and the transmittance of lucite. We found a good agreement between the observed and expected number of photoelectrons for both pixel and sum readouts. This suggests that MAROC sum electronics working as expected. Then we added background using LED up to 320 kHz/pixel, similar to SoLID running conditions. We can observe how the signal and background behave as shown in Figure~\ref{pixel_accuracy_cosmic}. Because of the relatively large signal from cosmic muons, the background can be separated using simple number of photoelectron cuts to reach better than 90\% accuracy for all 3 background levels. The pattern recognition using Hough transformation is applicable to the cosmic test because of the variation of the cosmic hit position, the partial ring and the lack of and optical focusing.

\subsection{Test summary}

We built and tested the MAROC sum readout system on bench and using LED and cosmic rays. These tests have shown that the electronics work well for the high rate requirement of the SoLID Cherenkov detectors. By providing analog sum of charge collected in pixels and pixel TDC information simultaneously, the MAROC sum board can be one of the potential options to read MAPMT photosensors in the SoLID Cherenkov detectors.

%% file: aiml.tex
\section{HGC PID using AIML}

\subsection{Introduction}

Machine learning (ML), a subfield of artificial intelligence (AI), has made tremendous progress over the last decade. In particular, deep-learning-based neural network models that use multiple layers with a large number of tunable parameters have demonstrated remarkable power in learning complex tasks. With recent advances in computing hardware, these models can leverage large volumes of data. Recent overviews of the use of AIML techniques in nuclear physics are available in two review articles~\cite{Bedaque2021,RevModPhys.94.031003}. They provide an up-to-date summary of work in the field and cover topics in nuclear theory, experimental methods, accelerator technology, and nuclear data.

ML has three main categories: supervised learning, unsupervised learning, and reinforcement learning.
Supervised learning involves learning from data that have associated labels, whereas unsupervised learning identifies patterns in datasets without requiring labels.
In nuclear physics applications, supervised learning models are often developed and trained on simulated data because simulated data come with labels.
Once the models are trained on simulated data, they are either applied directly to real data or fine-tuned beforehand using a smaller amount of labeled real data.
Unsupervised learning is also used in a number of nuclear physics applications for clustering and pattern discovery.

Supervised learning models can be further categorized as either classification or regression models.
In general, nuclear particle identification detectors can be treated as classification problems because the goal is to build a model that correctly assigns particles to a particular type or class.



Previous studies using AIML for Cherenkov type detectors were mostly focused on Ring-imaging Cherenkov (RICH) detectors and Detectors of Internally Reflected Cherenkov light (DIRC). They use very small photon sensors to collect Cherenkov photon rings and use their locations and sizes to distinguish charged particles such as electrons, pions, kaons, and protons over a wide momentum range. The particle type can be identified by classifying the corresponding detected hit pattern~\cite{fanelli2020machine}. Both the EIC ePIC dual-radiator RICH (dRICH)~\cite{Cisbani_2020} and the proximity-focusing RICH (pfRICH)~\cite{ML_pfRICH} use ML to improve PID. For the GlueX DIRC, DeepRICH~\cite{fanelli2020deeprich} was recently developed as a custom architecture that combines variational autoencoders, convolutional neural networks, and artificial neural networks. Implemented on GPUs, it improves reconstruction speed while maintaining accuracy close to that of the established FastDIRC method~\cite{hardin2016fastdirc} by enabling parallel processing of particle batches during inference.

Threshold Cherenkov detectors are traditionally equipped with large photon sensors that count photons produced by Cherenkov light from signal particles, while largely ignoring the spatial information carried by the Cherenkov rings. In the high-rate, high-background environment of SoLID, the threshold Cherenkov detectors LGC and HGC employ a unique readout system that provides detailed photon-position information. This ring information can then be used to help reject background and enhance performance. HGC PID is more challenging because it has higher background from lower Cherenkov threshold and its further downstream location. 

\subsection{Simulation and NPE method}

SoLID's Geant4 simulation includes all subdetectors and the three-dimensional magnetic-field map, as shown in Figure~\ref{fig:setup}. The HGC simulation includes Cherenkov processes and the wavelength-dependent optical properties of the gas, mirror, light collection cone, and PMT quantum efficiency. This level of detail is important for producing realistic signal and background simulations matching actual running conditions.

Each 5-cm-wide MAPMT in the HGC consists of 64 pixels (6$\times$6~mm$^2$). Our MAROC sum readout system enables simultaneous readout of individual pixels, quad sums of 16 pixels, and the total sum of 64 pixels. In one sector, 16 MAPMTs provide 1024 pixel channels, 64 quad channels, and 16 PMT signal channels. When a pion with momentum above 2.5~GeV/$c$ enters the HGC, its Cherenkov ring reflects from the mirrors and light-collection cones and produces photoelectrons in roughly two or three neighboring sectors. By contrast, a kaon with momentum below 7.5~GeV/$c$ does not emit Cherenkov light, allowing for $\pi/K$ separation. However, it can still produce secondary electrons through interactions with the thin aluminum window and with materials such as air and the detectors located upstream of the HGC.

\begin{figure}[hbt!]
\vspace*{-0.1in}
\centering
\includegraphics[width=0.7\textwidth]{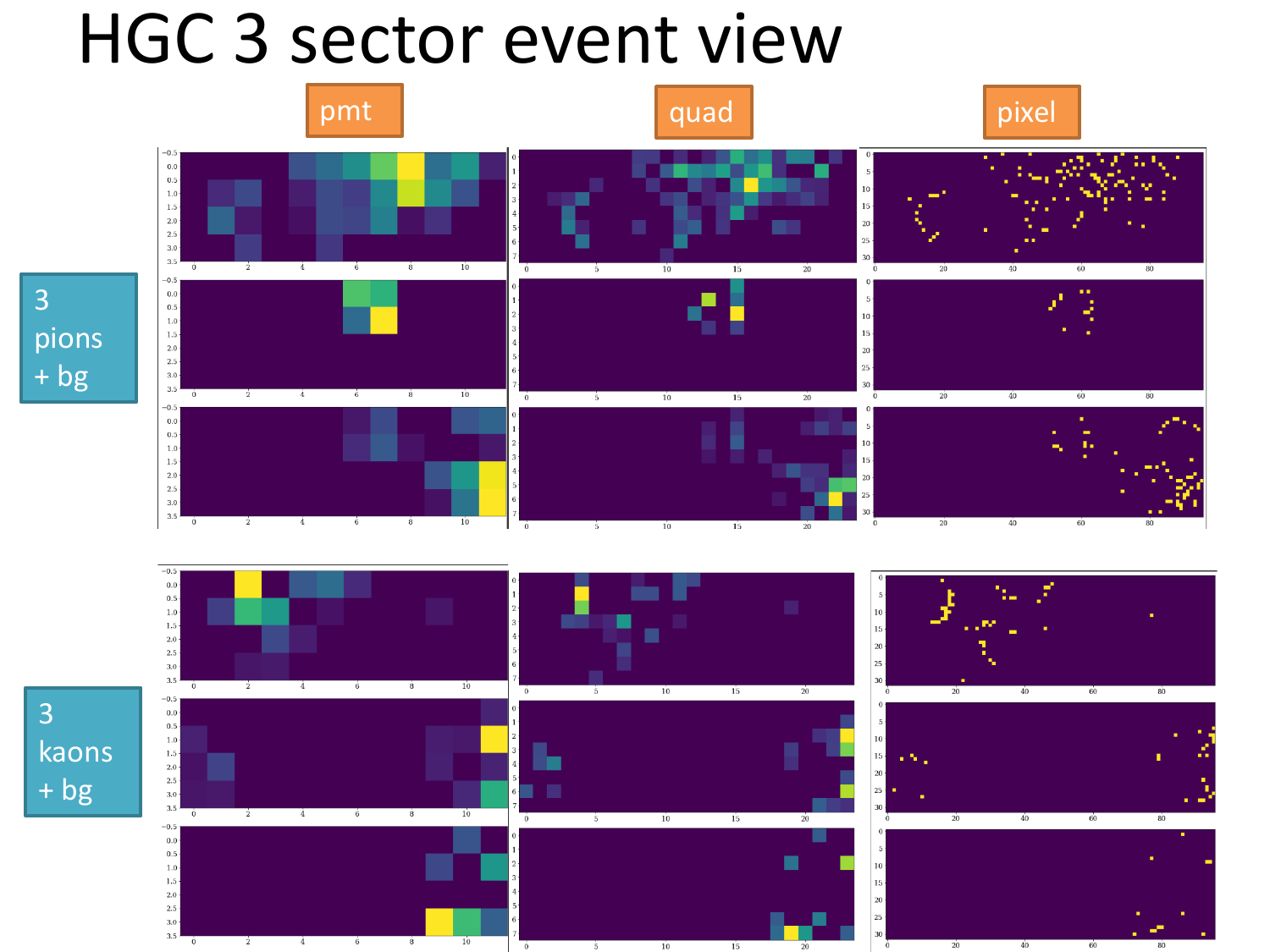}
\caption{Three sample pion and three sample kaon HGC events with beam background in three neighboring HGC sectors, obtained from the SoLID Geant4 simulation. From left to right, the columns show the same events at different spatial resolutions using PMT, quad, and pixel readout. The color scale indicates the number of photoelectrons, with 0 shown in dark purple. the pixel plots have only 1 or 0 values because the MAROC pixel readout works under a threshold Time-to-Digital Converter (TDC) mode. A fixed kinematic point is shown for pions and kaons with momentum 2.5~GeV/$c$, polar angle 8$^\circ$, and an arbitrary azimuthal angle from the target center.}
\label{fig:hgc_event}
\end{figure}

In fact, any low-energy electron with momentum greater than 8~MeV/$c$ can also produce background Cherenkov photons in the HGC. In the SoLID SIDIS $^3$He configuration at a luminosity of 10$^{37}$~cm$^{-2}$s$^{-1}$, the background rate from such electrons, produced by the 11~GeV electron beam interacting with the 40-cm-long pressurized $^3$He gas target, is about 200~kHz/pixel, or 4~MHz/PMT. These beam-produced background Cherenkov photons are mixed with simulated pion and kaon events within a readout time window of about 50~ns to mimic real data. Figure~\ref{fig:hgc_event} shows three sample pion events and three sample kaon events with beam background in three neighboring HGC sectors. Visually it is very hard to distinguish those two types of pion and kaon events, but a robust HGC PID algorithm must distinguish them in the presence of background with high confidence.

\begin{figure}[hbt!]
\vspace*{-0.1in}
\centering
\includegraphics[width=0.48\textwidth]{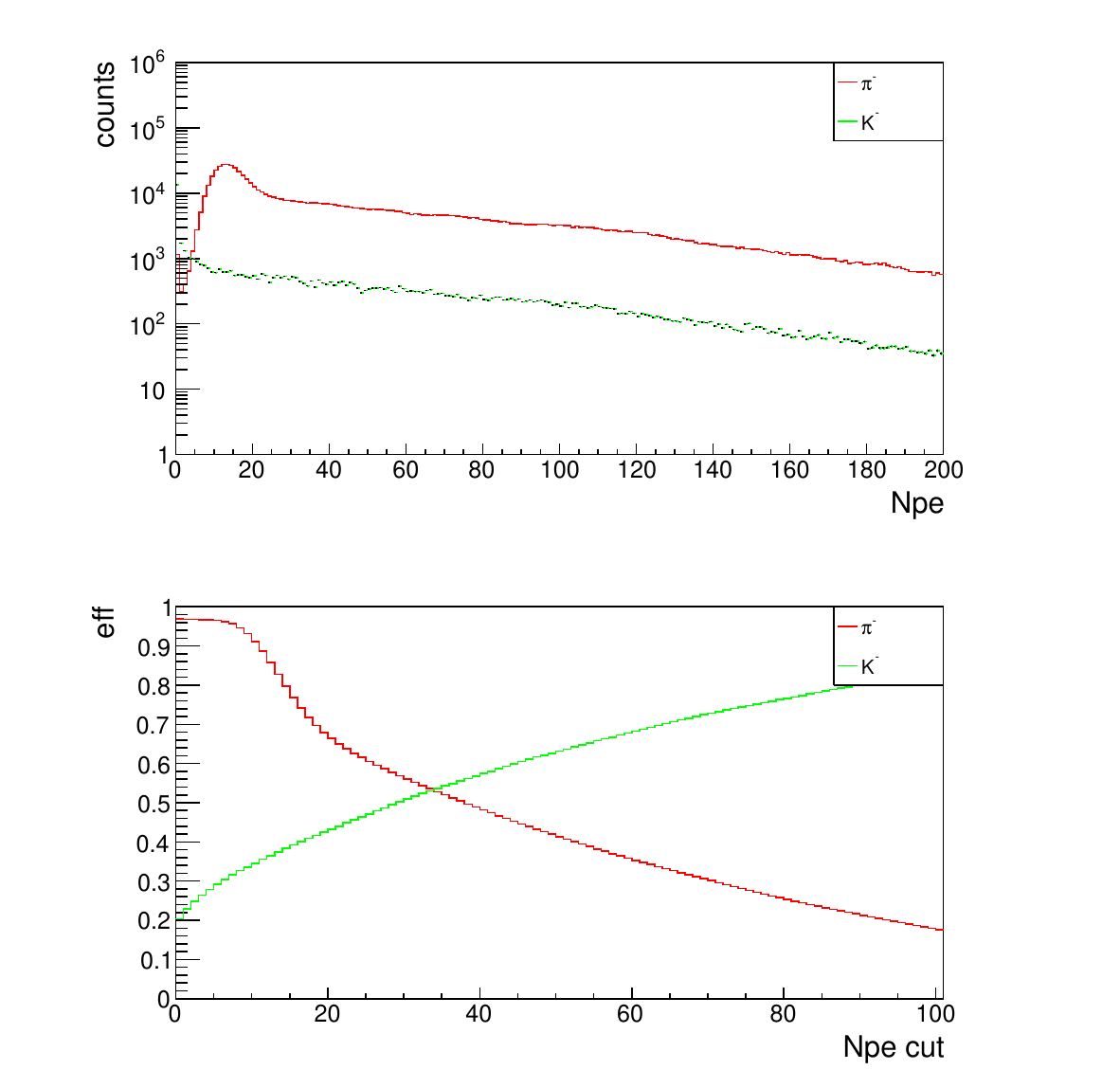}
\includegraphics[width=0.48\textwidth]{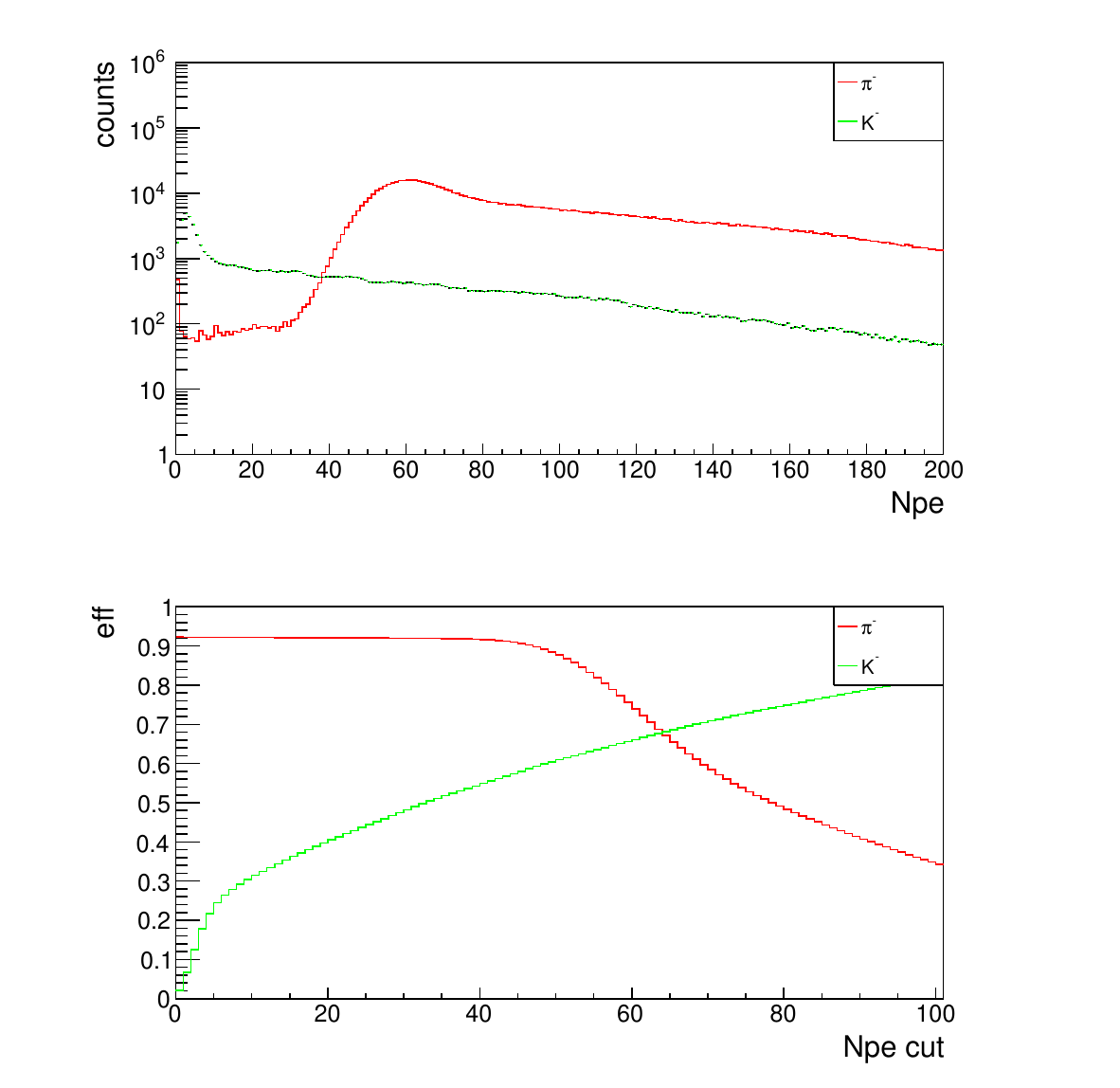}
\caption{HGC pion and kaon NPE counts (top) and NPE cut efficiency (bottom) at momentum 2.5~GeV/$c$ and polar angle 8$^\circ$ (left) and momentum 7.5~GeV/$c$ and polar angle 14.5$^\circ$ (right).}
\label{fig:hgc_npe_cut}
\end{figure}

The kinematic point at momentum 2.5~GeV/$c$ and polar angle 8$^\circ$ is the most challenging because pions at low momentum and small angle emit the fewest photons. The opposite extreme, corresponding to momentum 7.5~GeV/$c$ and polar angle 14.5$^\circ$, presents a different challenge when the background is high. The number of photoelectrons (NPE), including the effects of photon production and PMT quantum efficiency, for these two cases is shown in the top panels of Figure~\ref{fig:hgc_npe_cut}. Because we select only pion and kaon events that do not decay before entering the HGC after traveling about $\sim$7~m from the target, the ideal HGC performance would separate them with 100\% efficiency. A simple NPE cut can identify both types of events simultaneously and changing the cut value alters both efficiencies, as shown in the bottom panels of Figure~\ref{fig:hgc_npe_cut}. A balanced choice of cut value occurs at the intersection of the two efficiency curves, but it provides only 50--60\% efficiency for both pions and kaons. This performance is far below our target of above 90\% efficiency for both, which means HGC can identify 90\% of pion events while rejecting 90\% of kaon events.

A traditional likelihood-based pattern-recognition algorithm should perform better than the simple NPE cut, but it would require significant development effort and would be very challenging to optimize for such a high level of background. By contrast, AIML-based binary classification is well suited to this problem and has emerged as one of the most successful AIML applications.

\subsection{AIML method and results}

To prepare the HGC simulation data for AIML training, we stored them in a simple comma-separated values (CSV) format. Each row consists of a series of NPE entries for every readout unit in three neighboring sectors. For each kinematic setting, we generated three types of files for the same 1 million pion events and another three types for the same 1 million kaon events. For the PMT, quad, and pixel file types, there are 48, 192, and 3072 NPE values per row, respectively, as shown in Figure~\ref{fig:hgc_event}. We appended one final value as the event label, assigning 1 to pion events and 0 to kaon events.

We used a simple multilayer perceptron (MLP) neural network model for HGC PID training. The model consists of an input layer, five hidden dense layers, each with rectified linear unit (ReLU) activation and a 15\% dropout rate for improved generalization, and an output layer with sigmoid activation. The model was trained on 2 million labeled events for each kinematic setting and readout type. For each training, the data were divided into training, validation, and testing sets with a 50--25--25\% split.

\begin{figure}[hbt!]
\vspace*{-0.1in}
\centering
\includegraphics[width=0.45\textwidth]{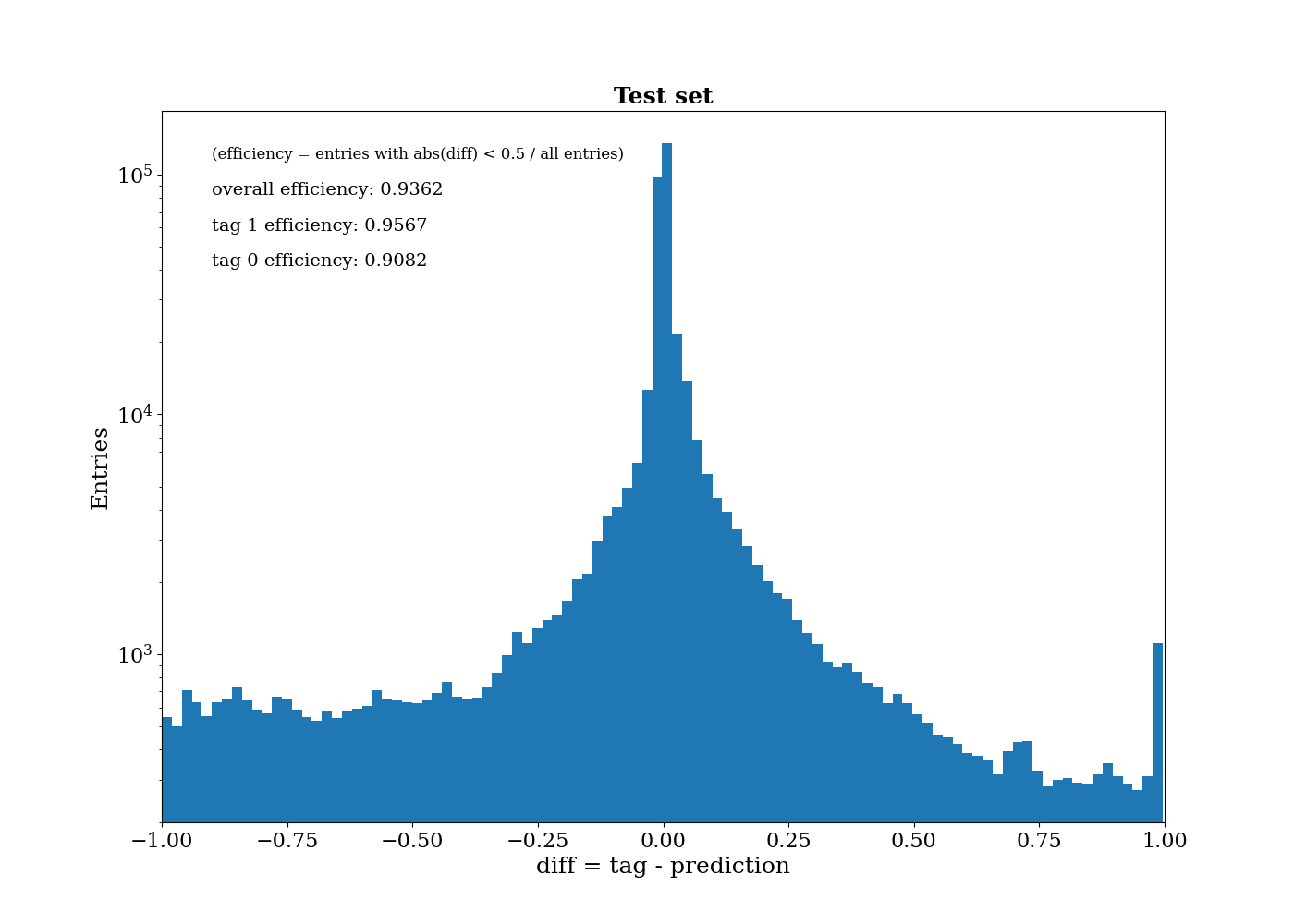}
\includegraphics[width=0.45\textwidth]{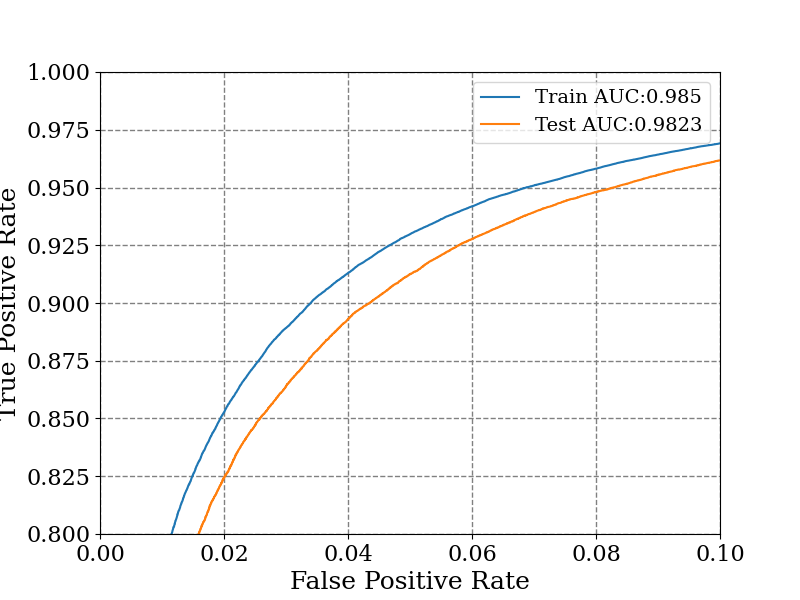}
\includegraphics[width=0.45\textwidth]{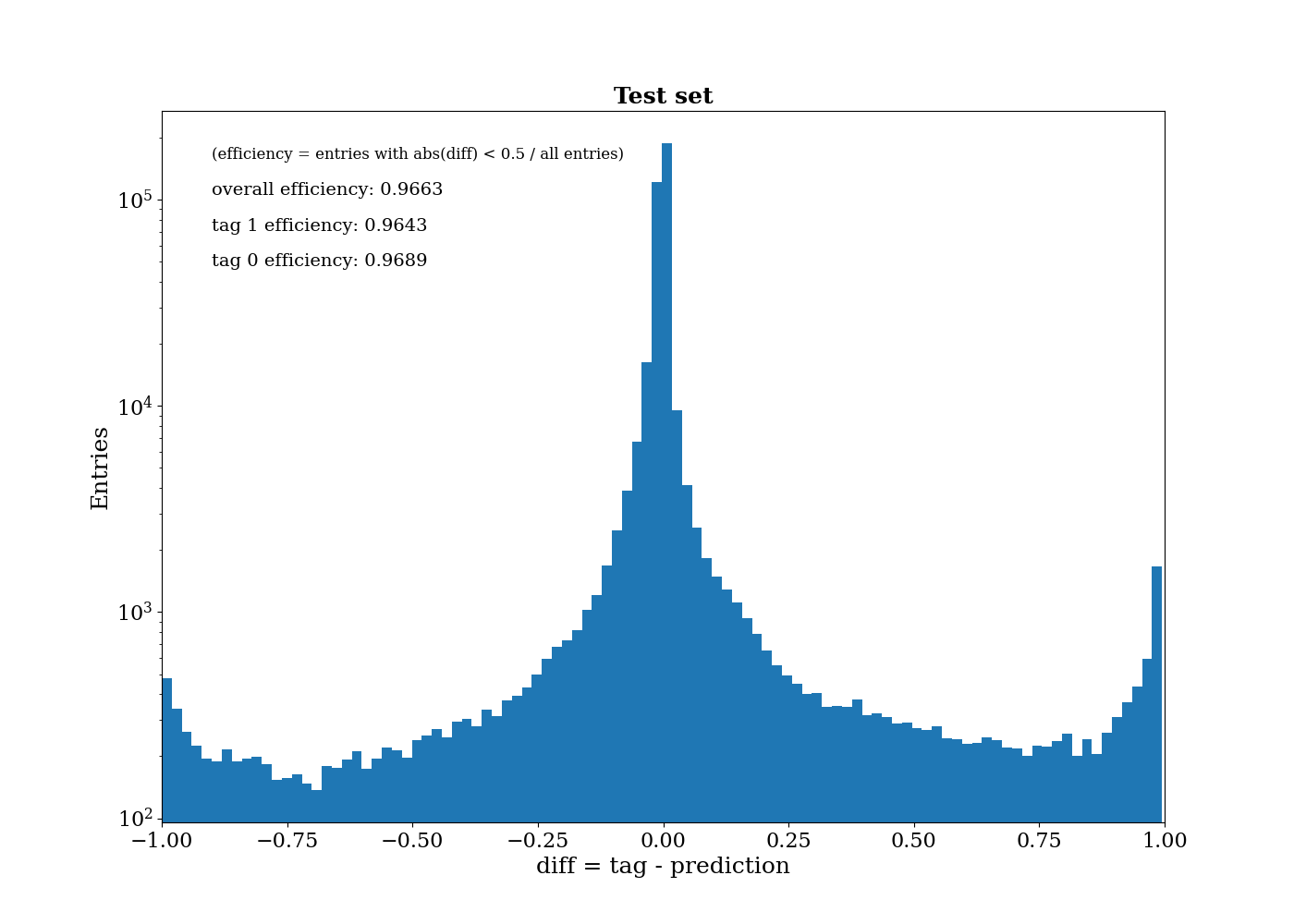}
\includegraphics[width=0.45\textwidth]{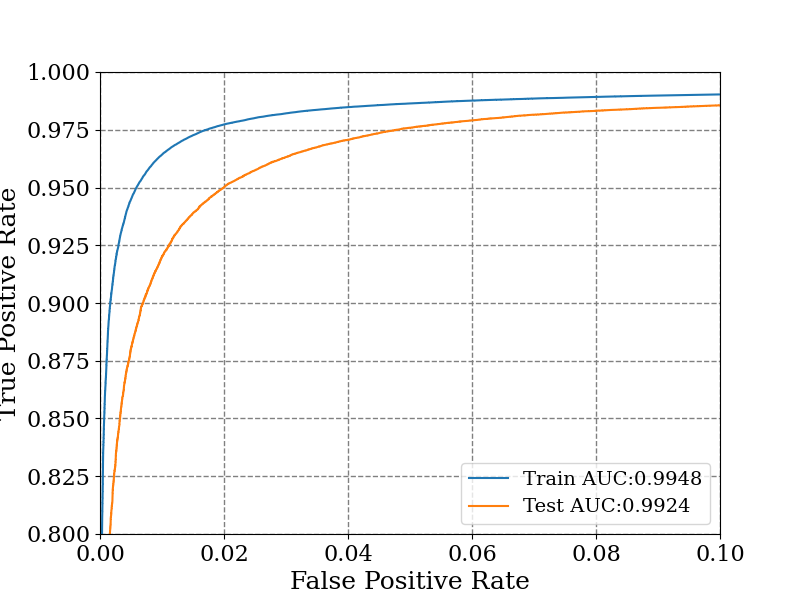}
\includegraphics[width=0.45\textwidth]{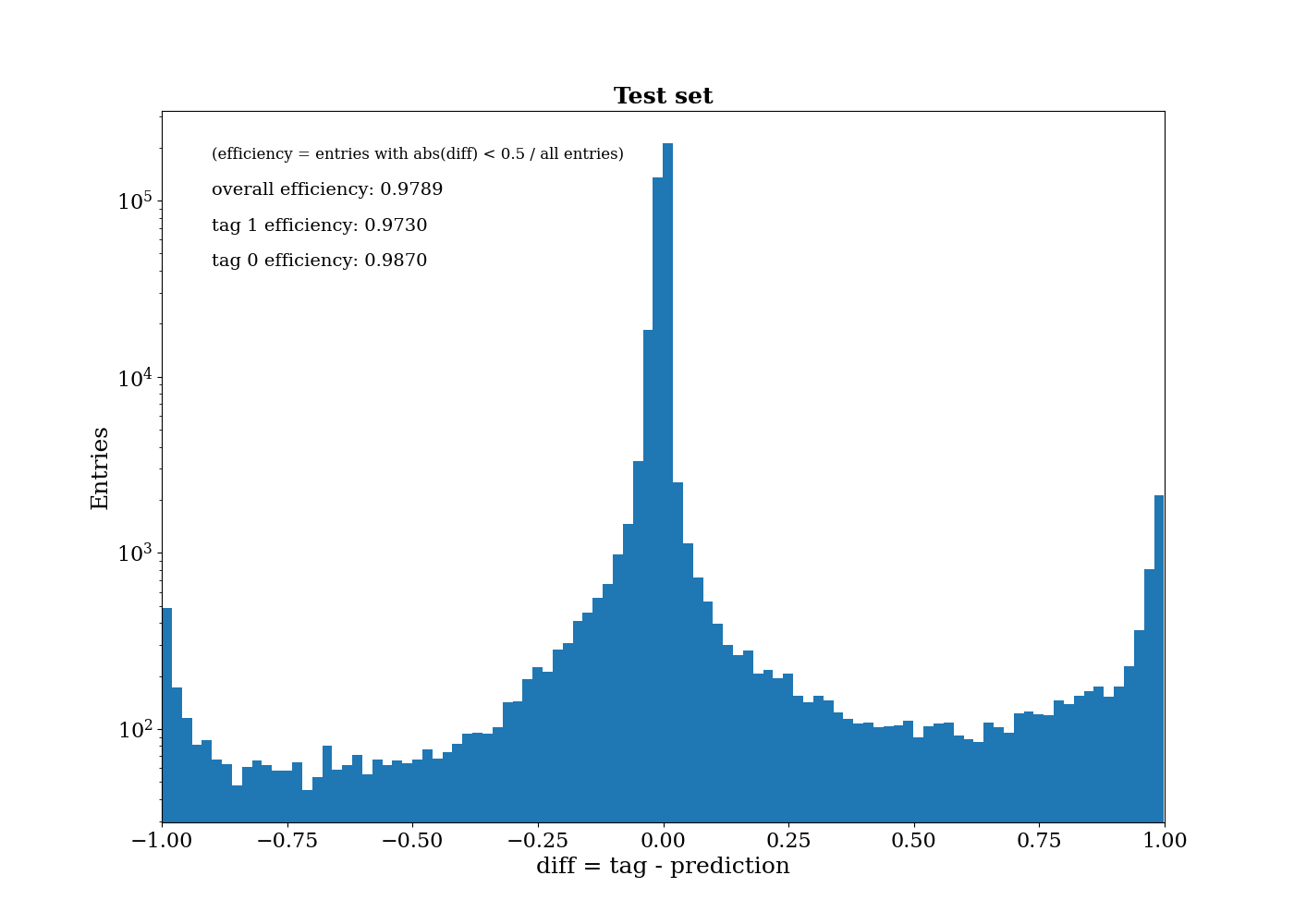}
\includegraphics[width=0.45\textwidth]{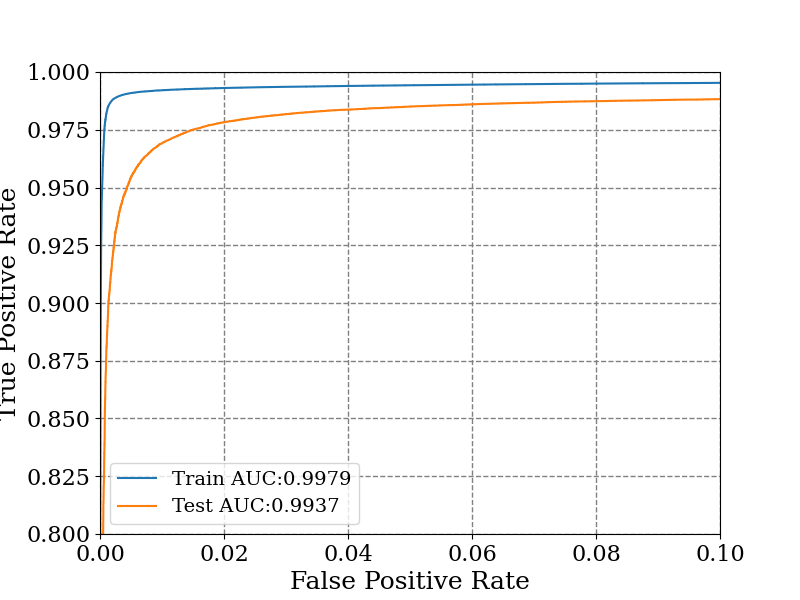}
\caption{Tag--prediction difference distributions (left) and AUC ROC (right) for PMT (top), quad (middle), and pixel (bottom) of HGC readout, obtained from models trained on simulation data at the kinematic point of 8$^\circ$ and 2.5~GeV/$c$.}
\label{fig:hgc_aiml}
\end{figure}

The results for all three readout schemes are shown in Fig.~\ref{fig:hgc_aiml} for the kinematic point of 2.5~GeV/$c$ and 8$^\circ$. The tag--prediction difference distributions for the testing sets, shown in the left panels, peak at 0 for correctly identified events. A balanced threshold choice of 0.5 yields both pion and kaon efficiencies on he plots. The right panels show the area under the receiver operating characteristic curve (AUC--ROC) of the training and testing sets, a standard performance metric for binary classification models that measures the ability to distinguish between positive and negative classes across a range of thresholds. Larger AUC values indicate better-performing models. Those plots show the PMT model performs somewhat worse, whereas the quad and pixel models have similar performance.

The overall results for the testing set are summarized in Table~\ref{tab:hgc_aiml}.
The low-NPE kinematic point at momentum 2.5~GeV/$c$ and polar angle 8$^\circ$ has the lowest efficiency and AUC, as expected. In particular, the 90.82\% kaon efficiency for the PMT case barely meets our target. The high-background kinematic point at momentum 7.5~GeV/$c$ and polar angle 14.5$^\circ$ turns out to be easier for all readout options. We also added a kinematic setting in which particles are distributed uniformly over the momentum range 2.5--7.5~GeV/$c$ and the polar-angle range 8--14.5$^\circ$. The corresponding results lie between those of the two extreme settings and suggest that a single broadly trained model could cover the full kinematic range, greatly simplifying future PID algorithm deployment. More generally, for all kinematic settings, the PMT readout option does not appear to provide sufficient information for PID, the quad option appears to provide enough information, and the pixel option yields only modest additional improvement.

The neural network model is not easy to interpret, and the machine-learning training and inference processes are fundamentally probabilistic rather than deterministic. To estimate the uncertainties in the AIML PID results, we explored several approaches, including running predictions multiple times with the trained models and repeating the same training process multiple times. We find that the results are stable at the $<1\%$ level. A more detailed study of these uncertainties will be an important topic for future work when we apply AIML PID to real data.


\begin{table}[!h]
\begin{center}
\scalebox{1.0}{
\begin{tabular}{|c|c|c|c|}
\hline 
eff $\pi/K$(AUC)  & PMT & quad & pixel\\
\hline
2.5GeV 8deg  & 0.9567/0.9082 (0.9823)  & 0.9643/0.9689 (0.9926) & 0.9730/0.9870 (0.9937)\\
\hline   
7.5GeV 14.5deg  & 0.9866/0.9871 (0.9981) & 0.9885/0.9950 (0.9987) & 0.9885/0.9943 (0.9982) \\
\hline
2.5-7.5GeV 8-14.5deg & 0.9673/0.9500 (0.9902) & 0.9722/0.9716 (0.9939) & 0.9755/0.9762 (0.9934) \\
\hline
\end{tabular}}
\end{center}
\caption{Efficiency for predicting $\pi/K$ using PMT, quad, and pixel HGC readout for models trained on simulation data at different kinematic points.}
\label{tab:hgc_aiml}
\end{table}


The AIML-based PID performance for HGC is greatly improved relative to that of the NPE-cut method, and the model is easier to train and implement than a traditional likelihood-based pattern-recognition algorithm. Our studies, together with the experience described in the previous section, provide a glimpse of AIML's potential and of how it can help SoLID Cherenkov detectors achieve their best performance under high-rate, high-background running conditions. The working software pipeline also suggests a practical path for carrying out these tasks. 



%% file: sum.tex
\section{Summary}


We present the development of readout electronics and AIML-based particle-identification methods for the SoLID Cherenkov detectors at Jefferson Lab, with emphasis on the heavy-gas Cherenkov detector (HGC). To preserve the spatial information of Cherenkov light patterns, we developed a MAROC sum readout system that provides simultaneous pixel, quadrant-sum, and total-sum signals from each MAPMT. Bench studies show that the system can operate at rates comparable to or above those expected for SoLID, with acceptable pedestal behavior and good signal linearity.

We also studied HGC particle identification using realistic Geant4 simulations with beam-related background. A simple photoelectron-counting cut is inadequate under these conditions, whereas multilayer perceptron models trained on PMT, quad, and pixel readout data perform substantially better. The quad and pixel readout schemes achieve pion and kaon efficiencies above 90\% and clearly outperform the PMT-only option.

Overall, the combination of high-rate MAROC sum electronics and AIML-based pattern recognition provides a practical path toward robust SoLID Cherenkov detector performance.